\documentclass[showpacs,aps,prb,twocolumn,groupedaddress,amsmath,amssymb]{revtex4}

\usepackage{graphicx}
\usepackage{color}
\usepackage{wrapfig}
\usepackage{soul}

\newcommand{\ev}[1]{\ensuremath{\langle #1 \rangle}}
\renewcommand{\Re}{\ensuremath{\operatorname{Re}}}

\newcommand{\abs}[1]{\ensuremath{|#1|}}

\newcommand{\p}{\ensuremath{^{\phantom{\dag}}}}   

\newcommand{\Eq}[1]{Eq.\,(\ref{#1})}

\newcommand{\Fig}[1]{Fig.\,\ref{#1}}
\newcommand{\Sec}[1]{Sec.\,\ref{#1}}

\begin{document}
\bibliographystyle{aip}

	\title{Transport through single-level systems: Spin dynamics in the nonadiabatic regime}
	
	\author{A. Metelmann}
	\email[]{metelmann@itp.tu-berlin.de}
	\author{T. Brandes}
	\affiliation{Institut f\"ur Theoretische Physik, TU Berlin, Hardenbergstr. 36, D-10623 Berlin, Germany}
	\date{\today}
	
\begin{abstract}
We investigate the Fano-Anderson model coupled to a large ensemble of spins under the influence of an external magnetic field.
The interaction between the two spin systems is treated within a meanfield-approach and we assume an anisotropic coupling between these two systems. By using a nonadiabatic approach, we make no further approximations in the theoretical description of our system, apart from the semiclassical treatment. Therewith, we can include the short-time dynamics as well as the broadening of the energy levels arising due to the coupling to the external electronic reservoirs. We study the spin dynamics in the regime of low and high bias. For the infinite bias case, we compare our results to those obtained from a simpler rate equation approach, where higher-order transitions are neglected. We show, that these higher-order terms are important in the range of low magnetic field. Additionally, we analyze extensively the finite bias regime with methods from nonlinear dynamics, and we discuss the possibility of switching of the large spin.
\end{abstract}


\maketitle

\section{Introduction}
A quantum dot typically consists of $10^5$ atoms. Electrons tunneling through such devices experience hyperfine and spin-orbit interaction with the nuclear spins of the host material.\cite{Hanson2007,Erlingsson2005,Baugh2007} Combinations of huge numbers of spins can be described as a large external effective spin system which interacts with the single electron spin. In experiments with quantum dots, features such as a large Overhauser field \cite{Baugh2007} and self-sustained current oscillations have been observed.\cite{Ono2004}

The effects of spin-orbit and hyperfine coupling is within the scope of recent research for various kinds of systems, such as self-assembled quantum dots, \cite{Takahashi2010,Hamaya2008} carbon nanotubes \cite{Hauptmann2008} or molecular magnets.\cite{Jo2006,Heersche2006,Bogani2008, Misiorny2009,Friedmann2010,Bode2012} The latter are promising candidates for spintronics.\cite{Thomas1996}

The interaction of electron spins with a large spin reveals interesting nonlinear effects, and the system is known to exhibit chaotic behavior.\cite{Feingold1983,Peres1984, Peres1984b,Robb1998}  These effects appear for a closed system with anisotropic coupling and an external magnetic field. L\'opez-Moniz and co-workers \cite{Lopez-Monis2012} discussed the coupling to two external leads, which were assumed to be polarized. Within a rate equation approach, they found that the chaotic behavior survives for small magnetic fields.

Using a nonadiabatic approach in this paper, we extend this approach to the finite bias regime, which is not accessible within the rate equation method. Furthermore, the rate equations method is also restricted to first-order transitions, which we also extend with our nonadiabatic approach. 

The nonadiabatic approach works well for systems whose quantum fluctuations are assumed to be small and that are coupled to a nonequilibrium environment. Here, we adopt this method (that we had extensively tested for nanoelectromechanical systems \cite{Metelmann2011}) to the more complex situation of a collective spin instead of an oscillator variable. This semiclassical description is suitable for these systems as long the number of spins, which build up the ensemble, is comparatively large.\cite{Schuetz2012,Mosshammer2012,Morrison2008,Morrison2008a,Ribeiro2008}
For a further analysis of correlations between the spin states, for instance to study entanglement of the ensemble and the single electron spin, a quantum description is certainly required. \cite{Rudner2010}
However, within a semiclassical description, it is possible to discuss the main spin dynamics. Even though the nuclear spin dynamics can be assumed to be slower than the electron spin dynamics, \cite{Hanson2007,Bode2012} we found that in certain parameter regimes an adiabatic approximation is not sufficient to describe the system's dynamics.

Our paper starts with a detailed description of the model and the adaptation of the nonadiabatic approach. We directly derive the rate equation approach from our nonadiabatic method, which is presented in \Sec{Sec.:BerlinMadrid}. 
Afterwards, we compare the results of both methods in the infinite bias regime. 
In \Sec{sec.:SDSfinitebias} we start the investigation of the finite bias regime with a dynamical analysis based on an adiabatic approach with Green's functions. Therewith we interpret the nonadiabatic results for the spin dynamics.
Within our conclusion in \Sec{Sec.Conclusion}, we discuss the advantages and disadvantages of the methods used here.

\section{Model} \label{sec.:model}
We assume that a vertical magnetic field $B_z$ is applied to a Fano-Anderson model.\cite{Anderson1961,Fano1961} The magnetic field leads to a Zeemann splitting of the electronic level $\varepsilon_d$ into two levels,\cite{Zeemann1897} corresponding to $\varepsilon_{\sigma} = \varepsilon_d \pm \frac{1}{2} B_z$, with $\sigma \in \uparrow, \downarrow$. Without further interactions, this solely leads to two spin-dependent current channels.
Only electrons with spin-up (-down) can tunnel through the upper (lower) energy level. These energy levels are broadened due to the coupling to the leads, and for small magnetic fields an overlap of both channels exists, but there is no communication arranged between the two energy levels, and spin-flips cannot occur. Here we enable transitions between the two energy levels with the help of a large external spin, which interacts with the electronic spin. 
\begin{figure}[t]
	\centering
		 \includegraphics*[width=0.9\linewidth]{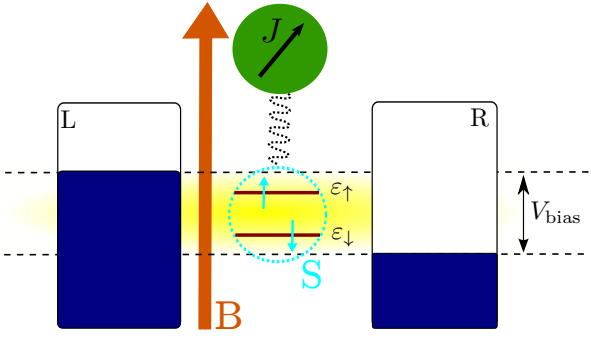}
  	 \caption{Sketch of a single-level system coupled to a large external spin $\textbf{J}$. Here $\textbf{S}$ denotes the electronic spin and B the external magnetic field. The detuning of the leads' chemical potentials leads to a transport window in the size of the applied bias $V_{\rm bias}$.}
	   \label{fic.:SDSsketch}
\end{figure}
\Fig{fic.:SDSsketch} depicts a sketch of the considered model. There, $\mathbf{\hat S}$ denotes the electronic spin operator for the levels, which components are defined via
\begin{align} \label{eq.:spinoperators}
\hat S_{x} =& \frac{1}{2}  \left(\hat d^{\dag}_{\uparrow} \hat d_{\downarrow}\p + \hat d^{\dag}_{\downarrow} \hat d_{\uparrow}\p \right), \nonumber \\
\hat S_{y} =& \frac{1}{2i} \left(\hat d^{\dag}_{\uparrow} \hat d_{\downarrow}\p - \hat d^{\dag}_{\downarrow} \hat d_{\uparrow}\p \right) , \nonumber \\
\hat S_{z} =& \frac{1}{2}  \left(\hat d^{\dag}_{\uparrow} \hat d_{\uparrow}\p   - \hat d^{\dag}_{\downarrow} \hat d_{\downarrow}\p \right),
\end{align}
introducing the creation/annihilation operators  $\hat d^{\dag}_{\sigma}/ \hat d_{\sigma}\p $ of the electronic levels.
The vertical magnetic field $B_z$ couples to the $\hat S_z$ operator, leading to the splitting of the initial single level. 
The Hamiltonian for the Fano-Anderson model reads
\begin{align} \label{eq.:FanoAndersonSpin}
 \mathcal H_{\rm{FA}} =& \sum_{\sigma} \varepsilon_{d} \ \hat d^{\dag}_{\sigma} \hat d_{\sigma}\p 
                        + \sum_{k \alpha \sigma}  \varepsilon_{k \alpha \sigma} \hat c_{k \alpha \sigma}^{\dag} \hat c_{k \alpha \sigma}\p  
                        \nonumber \\ &
                        + \sum_{k \alpha \sigma} \left( V_{k \alpha \sigma}  \hat c_{k \alpha \sigma}^{\dag} \hat d_{\sigma}\p
                        +  V^{\ast}_{k \alpha \sigma} \hat d^{\dag}_{\sigma} \hat c_{k \alpha \sigma}\p \right) 
                        +  B_{z} \hat S_{z}.
\end{align}
Note, that the left ($\alpha = \rm L$) and right ($\alpha = \rm R$) lead operators $\hat c^{\dag}_{k\alpha \sigma}/ \hat c_{k\alpha \sigma}\p $ are spin-dependent. This enables us to consider polarized leads, where the density of states for spin-up and spin-down electrons with energy $\varepsilon_{k \alpha \sigma}$ differs. This could be realized with ferromagnetic leads,\cite{Hamaya2008,Hofstetter2010,Sothmann2010,Weymann2005} leading to a spin-dependent current through the system.
The third term in the Hamiltonian describes the transitions between a state in the lead and the electronic levels with tunneling amplitude $V_{k \alpha \sigma}$. For simplicity, we include the prefactor of the last term in \Eq{eq.:FanoAndersonSpin}, containing the electronic $g$-factor, into the definition of the magnetic field.

In addition to the electronic spin operators, the large external spin's $z$-component $\hat J_z$ couples to the magnetic field. The free motion for it is described by
\begin{equation} \label{eq.:HJ}
 \mathcal H_{J}= B_z \hat J_z,
\end{equation}
where we assume the same g-factor and therewith the same magnetic field as for the electronic spin operators. This is a simplification in theoretical approaches, cf. \cite{Lopez-Monis2012, Bode2012}, but a generalization is straightforward.

Here, a large external spin means an effective spin describing a big ensemble of spins, for example, the collective spin of the nuclei in a quantum dot or a molecule. Electrons which tunnel through such devices, experience an interaction with the effective spin of the whole system. This interaction is described by \cite{Slicher1990,Hanson2007}
\begin{equation} \label{eq.:SpinSpinInteraction}
\hat V = \sum_{i} \lambda_i  \ \hat S_i   \hat J_i , \hspace{1cm} i = x,y,z;
\end{equation}
introducing the coupling constant $\lambda_i$. If these coupling constants are equal for all components $\lambda_i=\lambda_j$, one speaks of an isotropic coupling, corresponding to the Fermi contact term in the hyperfine interaction. The latter is important for spin-spin interactions in quantum dots, i.e. in GaAs dots due to their s-type conduction band.\cite{Cerletti2005} In a realistic quantum dot model the electron wave function is not uniform over the nuclei side and the coupling between the individual spins is varying, but it can be assumed to be piecewise flat in a large spin model.\cite{Rudner2010}

In this paper we consider the anisotropic case in which interesting dynamical behavior was observed.\cite{Robb1998}
There, at least for two components, $\lambda_i\neq\lambda_j$ is valid. An anisotropic coupling is relevant for systems with higher angular momentum bands where an enhancement of the anisotropic hyperfine interaction appears and the isotropic interaction vanishes.\cite{Coish2009} The anisotropic hyperfine interaction appears, for example, in carbon nanotubes or graphene,\cite{Fischer2009} as well as in molecular magnets.

We treat the interaction of the two spins in a semiclassical manner. Therefore, we have to assume that the quantum fluctuations in the system are small. This should be valid as long as the external spin is large and its fluctuation can be neglected. As a consequence of this assumption, we have no spin decay due to dissipation, and the large spin's length $j$ is conserved.

Using a mean-field approximation \cite{Lopez-Monis2012} for \Eq{eq.:SpinSpinInteraction} we obtain
\begin{equation}
\hat V_{\rm{MF}} = \sum_{i} \lambda_i \left( \hat S_i  \ \ev{\hat J_i} +  \hat J_i  \ \ev{\hat S_i} - \ev{\hat S_i}  \ \ev{\hat J_i} \right) , \hspace{0cm} i = x,y,z.
\end{equation}
Thereby, the fluctuations $ \delta \hat A_i = \hat A_i - \ev{\hat A_i} , \hat A \in \hat S, \hat J$ have been neglected. Now we can build up a closed system of equations for the considered system. 

For the large spin we use the commutation relations to derive the Heisenberg equations of motion
\begin{align} \label{eq.:LargeSpinEOM}
 \frac{d}{dt}\ev{\hat J_x} &= -\left( \lambda_z  \ev{\hat S_z} + B_z \right) \ev{\hat J_y} + \lambda_y  \ev{\hat S_y}  \ev{\hat J_z}, \nonumber \\
 \frac{d}{dt}\ev{\hat J_y} &=  \hspace{0.4cm} \left( \lambda_z  \ev{\hat S_z} + B_z \right) \ev{\hat J_x} - \lambda_x  \ev{\hat S_x} \ev{\hat J_z}, \nonumber \\
 \frac{d}{dt} \ev{\hat J_z} &=  \hspace{0.5cm} \lambda_x \ev{\hat S_x} \ev{\hat J_y} - \lambda_y \ev{\hat S_y} \ev{\hat J_x}.
\end{align}
This is a strongly nonlinear system, since the values of the electronic spin components depend on the large spin. 

The treatment of the electronic spin plays a decisive role in the theoretical description of this system. By using a rate equation approach, the electronic spin components are obtained from equations of motions similar to \Eq{eq.:LargeSpinEOM}. Therewith, the short-time dynamics is included, but the contributions of the leads come in only as rates. Following from that, higher-order transitions are neglected and one is restricted to the infinite bias regime. 
One way of including higher-order transitions is by applying an adiabatic approximation, where the movement of the large spin is assumed to be slow compared to changes in the electronic subsystem. The electronic spin operators can then be derived via Keldysh Green's functions
\cite{keldysh1965,Kamenev2009,Haug2008} and are given in explicit expressions. The disadvantage of this method is that the short-time dynamics is missing.

In the next part of this paper we derive equations of motion for the electronic spin operators using a completely nonadiabatic approach. The latter enables us to include the short-time dynamics as well as higher-order transitions. Following from that, we can probe the rate equation and the adiabatic approach.

\subsection{Nonadiabatic approach} \label{sec.:SDSnoadiabatictheory}
In the framework of the nonadiabatic approach, we calculate all system quantities by considering their full time-dependence. In the following, we assume an anisotropic coupling $\lambda_y=0$ and $\lambda_x=\lambda_z=\lambda$, and together with the mean-field approach, we obtain an effective Hamiltonian for the electronic levels,
\begin{align} \label{eq.:SDShamiltonianEFF}
 \mathcal H_{\rm{c}}(t) \equiv & \sum_{\sigma} \varepsilon_{\sigma}(t) \ \hat d^{\dag}_{\sigma} \hat d_{\sigma}\p 
                        +  \frac{\lambda}{2} \ev{\hat J_x(t)} 
                            \left( \hat d^{\dag}_{\uparrow} \hat d_{\downarrow}\p + \hat d^{\dag}_{\downarrow} \hat d_{\uparrow}\p\right) 
                        + \mathcal H_{\rm T}, \nonumber \\ 
 \varepsilon_{\sigma}(t) \equiv & \ \varepsilon_d \pm \frac{B_z}{2} \pm \frac{\lambda}{2} \ev{\hat J_z (t)}, \hspace{0.5cm} \sigma = \uparrow, \downarrow.
\end{align}
 The term $\mathcal H_{\rm T}$ contains the coupling to the leads. Based on this effective Hamiltonian, we can describe the effects arising due to the coupling to a large external spin. The $\ev{\hat J_z(t)}$ - component solely leads to an additional shift of the electronic levels,\cite{Koenemann2012} but the coupling to the $\ev{\hat J_x(t)}$ - component enables transitions between both levels. This Hamiltonian corresponds to a two-level or a parallel double-dot system,\cite{Mourokh2002,Brandes2005}
 where the prefactor $\frac{\lambda}{2} \ev{\hat J_x(t)}$ would be equivalent to a time-dependent tunneling amplitude between the two (dot) levels $\varepsilon_{\uparrow} $ and $\varepsilon_{\downarrow} $.

The derivation of the equations of motion for the spin operators is performed starting from the Heisenberg equations of motion. 
The derivation is similar to those used before for the description of nanoelectromechanical systems, for details see Ref.\cite{Metelmann2011}. 
The probability for a transition between lead $\alpha$ and the electronic level $ \varepsilon_{\sigma}$ for an electron with energy $\omega (\hbar \equiv 1)$ and spin $\sigma$ is described by the spin dependent tunneling rates $\Gamma_{\alpha \sigma} (\omega)$ obtained from Fermi's Golden rule. We use a flat band approximation leading to energy-independent rates $\Gamma_{\alpha \sigma} = 2 \pi \sum_k \left|V_{k \alpha \sigma}\right|^{2} \delta (\omega -\varepsilon_{k \alpha \sigma})$.
Finally, the results for the spin operator expectation values yield $(\Gamma \equiv \Gamma_{\sigma}= \sum_{\alpha} \Gamma_{\alpha \sigma}$)
\begin{align} \label{eq.:SpinOperatorsEOM}
\frac{d}{dt} \ev{\hat S_x(t)}  =& - \Gamma \ev{\hat S_x(t)} 
                               - \left(B_z + \lambda \ev{\hat J_z(t)}\right) \ev{\hat S_y(t)} \nonumber \\ &
                               + \sum_{\alpha} \int d\omega \mbox{Re} \left[\mathcal T^{\alpha}_{\uparrow \downarrow}(\omega,t)
                               + \mathcal T^{\alpha}_{\downarrow \uparrow}(\omega,t) \right] , \nonumber\\
\frac{d}{dt} \ev{\hat S_y(t)}  =& - \Gamma \ev{\hat S_y(t)} 
                               + \left(B_z + \lambda \ev{\hat J_z(t)}\right) \ev{\hat S_x(t)} \nonumber \\ &
                               - \lambda \ev{\hat J_x(t)} \ev{\hat S_z(t)} \nonumber \\ &
                               + \sum_{\alpha} \int d\omega \mbox{Im} \left[\mathcal T^{\alpha}_{\uparrow \downarrow}(\omega,t)
                               - \mathcal T^{\alpha}_{\downarrow \uparrow}(\omega,t) \right] , \nonumber\\
\frac{d}{dt} \ev{\hat S_z(t)}  =& - \Gamma \ev{\hat S_z(t)} 
                               + \lambda \ev{\hat J_x(t)} \ev{\hat S_y(t)}   \nonumber \\ &
                               + \sum_{\alpha} \int d\omega \mbox{Re}  \left[ \mathcal T^{\alpha}_{\uparrow \uparrow}(\omega,t)
                              - \mathcal T^{\alpha}_{\downarrow \downarrow}(\omega,t) \right],
\end{align}
with the definition
\begin{align}
\mathcal T_{\sigma \sigma'}^{\alpha}(\omega,t) \equiv i \sum_k  V_{k \alpha \sigma} \delta(\omega-\varepsilon_{k \alpha \sigma})
e^{i \varepsilon_{k \alpha \sigma} t} \ev{\hat c^{\dag}_{k \alpha \sigma}(0) \hat d_{\sigma'}(t)},
\end{align}
for the lead-transition functions
\begin{align}\label{eq.:SDSleadtransitionfunctions}
\frac{d}{dt} \mathcal T^{\alpha}_{\sigma \sigma}(\omega,t)  =&
                              -i(\varepsilon_{\sigma}(t) - \omega - \frac{i}{2} \Gamma) \mathcal T^{\alpha}_{\sigma \sigma}(\omega,t)
                               \nonumber \\ &
                              - i \frac{\lambda}{2} \ev{\hat J_x(t)}  \mathcal T^{\alpha}_{\sigma \sigma'}(\omega,t)  
                              +  \frac{\Gamma_{\alpha \sigma}}{2\pi} f_{\alpha}(\omega) ,\nonumber\\
\frac{d}{dt} \mathcal T^{\alpha}_{\sigma \sigma'}(\omega,t) =&
                              -i(\varepsilon_{\sigma'}(t) - \omega - \frac{i}{2} \Gamma) \mathcal T^{\alpha}_{\sigma \sigma'}(\omega,t)
                              \nonumber \\ &
                              - i \frac{\lambda}{2} \ev{\hat J_x(t)}  \mathcal T^{\alpha}_{\sigma \sigma}(\omega,t),
\end{align}
where the function for equal spins $T^{\alpha}_{\sigma \sigma}$ couples to the one for different spins $T^{\alpha}_{\sigma \sigma'}$ and vice versa.
The time-dependent $z$-component of the large spin is included into the effective levels $\varepsilon_{\sigma}(t)$, cf. \Eq{eq.:SDShamiltonianEFF}. Here, the Fermi function $f_{\alpha}(\omega)$  does not depend on the spin due to the assumption, that the chemical potentials for spin-up and -down electrons in each lead are equal.

For the numerical calculations, we decompose the $\mathcal T^{\alpha}_{\sigma \sigma}$ into their real and imaginary part and discretize the integration over the lead energies in $N$ intervals. Therefore, we obtain $\left[ 6 + 16 (N+1) \right]$ coupled equations, including the equations of motion for the large spin, cf. \Eq{eq.:LargeSpinEOM}.

Using the definitions for the lead-transition functions \Eq{eq.:SDSleadtransitionfunctions}, the electronic current is obtained from
\begin{align}  \label{currentNonAd}
\mathcal I_{\alpha \sigma}(t) &=    e  \ \left\{  \int d\omega \ 2 \ \mbox{Re} \left[\mathcal T^{\alpha}_{\sigma \sigma} (\omega, t)\right] 
                                               - \Gamma_{\alpha\sigma} \ev{\hat n_{\sigma}(t)} \right\},
\end{align}
where  $e$ equals the electron charge and the equation of motion for the occupation yields
\begin{align} 
\frac{d}{dt} \ev{\hat n_{\sigma}(t)}=& - \Gamma \ev{\hat n_{\sigma}(t)} 
                               \pm \lambda \ev{\hat J_x(t)} \ev{\hat S_y(t)}   \nonumber \\ &
                               + \sum_{\alpha} \int d\omega \ 2 \ \mbox{Re}  \left[ \mathcal T^{\alpha}_{\sigma \sigma}(\omega,t)
                                \right].  
\end{align}
Here, the upper (lower) sign refers to $\uparrow (\downarrow)$ - electrons. Note, that the total occupation number of the dot $\ev{\hat N(t)} = \ev{\hat n_{\uparrow}(t)} +\ev{\hat n_{\downarrow}(t)}$ still couples to the spin operators via the transition functions. In the next section we derive a rate equation approach in which this dependence is omitted.

\subsection{Infinite bias: rate equation approach}\label{Sec.:BerlinMadrid}

We recover previous results for this model \cite{Lopez-Monis2012} by applying an adiabatic approximation to our approach. In concrete terms, we assume the large spin's movement as slow compared to the electrons which are entering the system. As a consequence, we neglect the time-dependence of the large spin in the equations for the lead-transition functions \Eq{eq.:SDSleadtransitionfunctions}, leading to $\varepsilon_{\sigma}(t)\equiv \varepsilon_{\sigma}$ and a decoupling from the equations for the large spin, $\ev{\hat J_x(t)} \equiv \ev{\hat J_x}$. The remaining two coupled equations can be solved via a Laplace transformation. Note, that this method is not a complete adiabatic approach, because in the end we still solve equations of motions for the electronic spin components, which is not the case for a full adiabatic ansatz.

Starting from \Eq{eq.:SDSleadtransitionfunctions} we obtain for $t\rightarrow\infty$
\begin{align} \label{eq.:SDSleadtransAdiabatic}
\mathcal T^{\alpha}_{\sigma \sigma}(\omega)  &= \frac{\Gamma_{\alpha \sigma}}{2 \pi} f_{\alpha}(\omega)
                              \frac{-i(\varepsilon_{\sigma'} - \omega - \frac{i}{2} \Gamma)}
                                   {N(\omega) },\nonumber\\
\mathcal T^{\alpha}_{\sigma \sigma'}(\omega) &=  \frac{\Gamma_{\alpha \sigma}}{2 \pi} f_{\alpha}(\omega)
                              \frac{i \frac{\lambda}{2} \ev{\hat J_x}}
                                   { N(\omega) }. 
\end{align}
with $N(\omega) = (\varepsilon_{\sigma'} - \omega - \frac{i}{2} \Gamma)(\varepsilon_{\sigma}  - \omega - \frac{i}{2} \Gamma) - \frac{\lambda^2}{4} \ev{\hat J_x}^2 $.
Before inserting these results into the equations for the spin operators \Eq{eq.:SpinOperatorsEOM}, we separate them into real and imaginary 
parts and perform the integrations over $\omega$.
Hence, the spin operator equations become
\begin{align} \label{eq.:SDSrateequations}
\frac{d}{dt} \ev{\hat S_x(t)}  =& - \Gamma \ev{\hat S_x(t)} 
                               - \left(B_z + \lambda \ev{\hat J_z(t)}\right) \ev{\hat S_y(t)}, \nonumber\\
\frac{d}{dt} \ev{\hat S_y(t)}  =& - \Gamma \ev{\hat S_y(t)} 
                               + \left(B_z + \lambda \ev{\hat J_z(t)}\right) \ev{\hat S_x(t)} \nonumber\\ &
                               - \lambda \ev{\hat J_x(t)} \ev{\hat S_z(t)}, \nonumber\\ 
\frac{d}{dt} \ev{\hat S_z(t)}  =& - \Gamma \ev{\hat S_z(t)} 
                               + \lambda \ev{\hat J_x(t)} \ev{\hat S_y(t)}  \nonumber\\ &
                               + \frac{1}{2} \left( \Gamma_{\rm L \uparrow} - \Gamma_{\rm L \downarrow} \right).
\end{align}
This result coincides with \cite{Lopez-Monis2012}: the whole system has now been reduced to six coupled equations, and the current simplifies to
\begin{align}  \label{eq.:currentRate}
 \mathcal I_{\alpha \sigma}(t) &=    e \Gamma_{\alpha \sigma}  \ \left\{ \delta_{\alpha \rm L}
                                               -  \ev{\hat n_{\sigma}(t)} \right\},
\end{align}
with the occupation
\begin{align} 
\frac{d}{dt} \ev{\hat n_{\sigma}(t)}=& - \Gamma \ev{\hat n_{\sigma}(t)} 
                               \pm \lambda \ev{\hat J_x(t)} \ev{\hat S_y(t)}  
                               + \Gamma_{\rm L \sigma}.  
\end{align}
As mentioned before, the total occupation number of the dot, $\ev{\hat N(t)} = \sum_{\sigma}\ev{\hat n_{\sigma}(t)}$, 
decouples from the remaining equations and becomes constant in the long-time limit,
\begin{align} \label{eq.:SDStotaloccupationRate}
 \ev{\hat N(t\rightarrow\infty)} =& \frac{\left(\Gamma_{\rm L \uparrow} +\Gamma_{\rm L \downarrow}\right)}{\Gamma}.
\end{align}
The equations for the spin operators \Eq{eq.:SDSrateequations} and the large external spin \Eq{eq.:LargeSpinEOM} provide further possibilities for an analytic investigation. The fixed points of the system can easily be calculated, see \cite{Lopez-Monis2012}. 

\begin{figure}
  \begin{center}
    \includegraphics[width=0.45\textwidth]{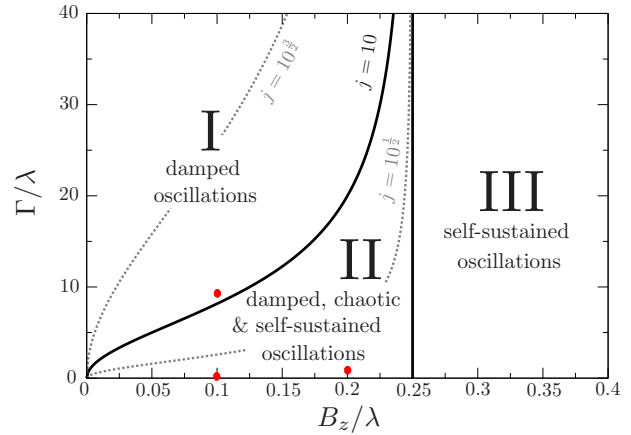}
  \end{center}
 \caption{\label{fig.:FixpunkteIB} Regions of different dynamical behavior concerning the parameters $\Gamma$ and $B_z$ using the rate equation approach. As a complement, the gray lines in \Fig{fig.:FixpunkteIB} illustrate the change of the first two regions if the length of the large spin is varied, while the ratio of the tunneling rates is fixed. For a smaller spin ($j=10^{\frac{1}{2}}$), region II decreases to the benefit of region I, which increases. Equally, the reverse situation can be seen, if the length of the spin is increased ($j=10^{\frac{3}{2}}$). Modifying the ratio of the tunneling rates shifts the vertical line separating region II from region III.  The red dots denote the parameter sets for region I and II which we used in our discussion below, see \Sec{Sec.:IBresults}. }
\end{figure}

By varying the magnetic field $B_z$ and the tunneling rate $\Gamma$ three regions of different dynamical behavior were obtained from the rate equation approach. In \Fig{fig.:FixpunkteIB} these regions are depicted.
There, the boundary between the regions is defined via the fixed points of the system, which we introduce below. The solid black lines correspond to parameters used in. \cite{Lopez-Monis2012} There, the length of the large spin equals $j = \abs{\hat J} = 10$ and the tunneling rates match $2\Gamma_{\rm L \uparrow} = \Gamma_{\rm L \downarrow} = \Gamma$. This choice of tunneling rates implies $\Gamma_{\rm R \downarrow} = 0$, because in this case the current flows exclusively through the upper electronic level. Following from that, spin-down electrons get trapped into the lower level and only contribute to transport after a spin-flip. Note, that $j=10$ is a relatively small value for a large spin in the semiclassical regime. A higher value of $j$  modifies the border between region I and region II, as depicted in \Fig{fig.:FixpunkteIB}, but the respective dynamical behavior in these regions stays qualitatively the same.

\begin{figure}[t]
	\centering
		 \includegraphics*[width=1.0\linewidth]{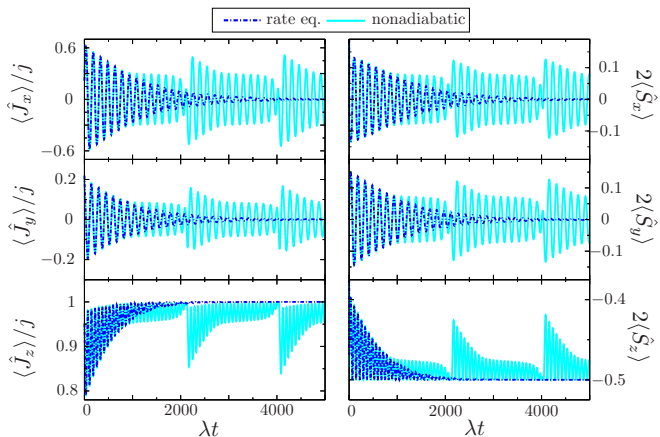}
  	 \caption{Comparison of the nonadiabatic approach and the rate equation approach results for regime I. The magnetic field equals $B_z/\lambda = 0.1$ and the tunneling rate is chosen as $\Gamma/\lambda = 9$. Initial conditions are $\ev{\hat J_x(0)} = \ev{\hat J_y(0)} = 5(\sqrt{5}-1)/(2\sqrt{2})$ and $\ev{\hat J_z(0)} = 5/\sqrt{2}\sqrt{5+\sqrt{5}}$ for the large spin and $\ev{\hat S_y(0)} = \ev{\hat S_z(0)} = 0$ and $\ev{\hat S_x(0)} = 0.5$ for the electronic spin.}
	   \label{RegimeINonAd}
\end{figure}

\subsection{Infinite bias (IB): results}\label{Sec.:IBresults}

The results for regime I are depicted in \Fig{RegimeINonAd}. The rate equation results describe damped oscillations, as expected. For large times, the $z$-component of the large spin becomes polarized parallel to the magnetic field and one spin-down electron gets trapped in the lower energy level. There the spin trajectories end up in one of the fixed points, 
\begin{align} \label{eq.:SDSfirstFIX}
\mathcal P^{\pm,\rm IB}_{S0}:
 \ev{\hat J_0} = \left(0,0,\pm j \right), \hspace{0.2cm}
 \ev{\hat S_0} = \left(0,0,\frac{\Gamma_{\rm L \uparrow} -\Gamma_{\rm L \downarrow}}{2 \Gamma} \right).
\end{align}
These stationary solutions exist in the whole parameter regime and are independent of the magnetic field.

In the nonadiabatic case the dynamical behavior is quite different. For small times, the spin trajectories follow the damped results from the rate equation approach. But the damping decreases strongly after the time step $t\lambda \approx 1000$. The same amount of time steps later, the amplitude drops down for one oscillation period, followed by a return to the initial oscillation. This behavior is similar for all spin components. If we consider the $z$-component for the large spin, the turning point appears when it approaches its fixed point value $\ev{\hat J_{z,0}} = j$. 

\begin{figure}[ht]
	\centering
		 \includegraphics*[width=1.0\linewidth,clip]{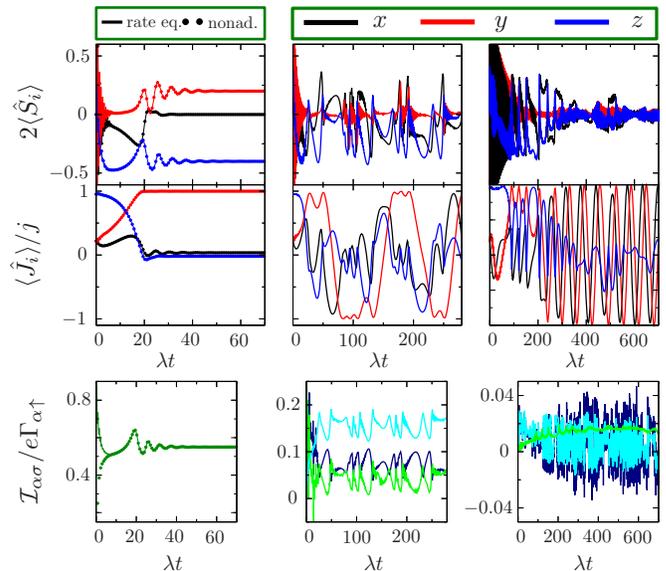}
  	 \caption{Behavior of the spin components in regime II. The left column depicts results for $B_{z}/\lambda = 0.2$ and $\Gamma/\lambda = 0.7$, where fast damping of the trajectories appears in both approaches. 
  	 The middle column shows results for $B_z/\lambda = 0.1$ and $\Gamma/\lambda = 0.16$, where the rate equations predict self-sustained oscillations, but for the nonadiabatic results, 
  	 already chaotic-like behavior appears. This is the same for the right column, where the tunneling rate is further reduced ($\Gamma/\lambda = 0.015$).  For clarity, we omitted the rate equation results for the chaotic-like regions. The lowest row depicts the corresponding current results. 
  	 For the chaotic-like
  	 regions, the green line denotes $ - \mathcal I_{\rm R \uparrow}$, the light blue line $\mathcal I_{\rm L \uparrow}$, and the dark blue line $\mathcal I_{\rm L\downarrow}$. Initial conditions are chosen as for \Fig{RegimeINonAd}.}
	   \label{RegimeIINonAd}
\end{figure}

By varying the parameters, we found no damped oscillations in region I at all. 
In the area close to the boundary between region I and region II, the dynamics appears as in \Fig{RegimeINonAd}. The time-interval between the
turning points decreases by going away from the border and deeper into region I. By keeping the magnetic field fixed at $B_z/\lambda = 0.1$ and increasing the tunneling rate, the trajectories perform smooth self-sustained oscillations. This transition appears near $\Gamma/\lambda \approx 11$. If we decrease the tunneling rate, the oscillations disappear around $\Gamma/\lambda \approx 8$ and we enter region II. Note that
the borders between all regions defined within the rate equation approach coincide with those in the nonadiabatic approach,
because the appearing fixed points are the same.

The results for regime II are depicted in \Fig{RegimeIINonAd}. The rate equation approach predicts three kinds of dynamical behavior, namely strong damped, self-sustained, and chaotic oscillations. 

For $B_z/\lambda = 0.2$ and small tunneling rate $\Gamma/\lambda=0.7$ the system performs strong damped oscillations and runs into one of the fixed points $\mathcal P^{+,\rm IB}_{SN} (N=1,2)$,
\begin{align} \label{eq.:SDSsecondFIX}
 \ev{ \hat J_{1,2}} =& \left(  \frac{\Gamma}{B_z} \mathcal B_{1,2},  \sqrt{j^2-\frac{\Gamma^2}{B_z^2} \mathcal B^2_{1,2}-\frac{B_z^2}{\lambda^2}}, - \frac{B_z}{\lambda} \right) , \nonumber \\ 
 \ev{\hat S_{1,2}} =&  \left(0, \mathcal B_{1,2}, - \frac{B_z}{\lambda} \right),
\end{align}
\begin{figure}[ht]
	\centering
		 \includegraphics*[width=1.0\linewidth]{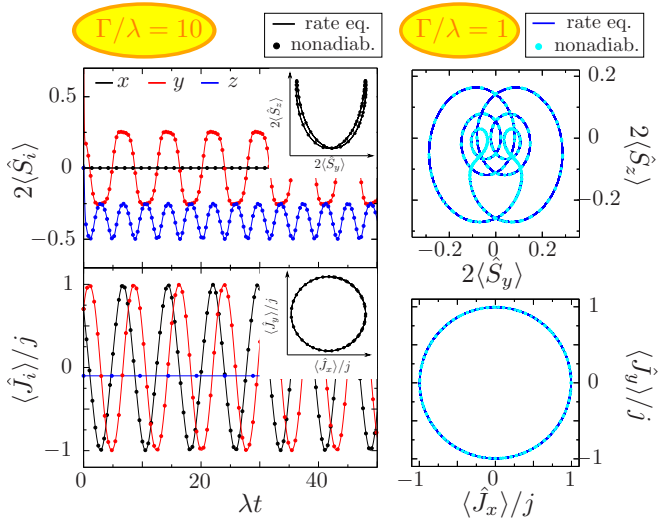}
  	 \caption{Results for regime III, where self-sustained oscillations appear. In the left graphs ($\Gamma/\lambda=10$) the oscillations are smoothed sinusoidal, the insets show the results in polar representation. For lower tunneling rate, the $\ev{\hat S_{i}}$ trajectories perform non-sinusoidal, but periodic oscillations, depicted in the upper graph on the right side. Parameters are $B_z/\lambda = 1.0 $, $\epsilon_d/\lambda = 0$ and with the initial conditions $\ev{\hat J_x(0)} = \ev{\hat J_y(0)} = 3\sqrt{5.5}$, $\ev{\hat J_z(0)} = -1$ and $\ev{\hat S_x(0)} = \ev{\hat S_z(0)} = 0$, $\ev{\hat S_y(0)} = 0.5$.}
	   \label{RegimeIIINonAd}
\end{figure}

\noindent with
\begin{align}
\mathcal B_{1,2} =& \pm \sqrt{\frac{B_z}{\lambda}  \left[\frac{\Gamma_{\rm L \downarrow} -\Gamma_{\rm L \uparrow}}{2 \Gamma} - \frac{B_z}{\lambda}\right]}.
\end{align}
Additionally, there exist two more fixed points $\mathcal P^{-,\rm IB}_{SN}$ with opposite sign of the $y-$component of the large spin.

This result coincides perfectly with the nonadiabatic result, see left column in \Fig{RegimeIINonAd}. The spin components run into the fixed point $\mathcal P^{+,\rm IB}_{S1}$, cf. \Eq{eq.:SDSsecondFIX}. The large spin becomes almost completely polarized perpendicular to the magnetic field.

In the first instance, by further increasing the tunneling rate, the rate equations forecast self-sustained oscillations for the systems
trajectories, followed by a chaotic oscillating behavior. Using the nonadiabatic approach, things change again. There, the chaotic-like behavior appears already in the parameter region, where self-sustained oscillations were predicted by the rate equation approach. Within the
nonadiabatic approach, we do not recover these self-sustained oscillations in region II; we solely observe strong damping or chaotic-like 
behavior.

For higher values of the external magnetic field (regime III) the system performs self-sustained oscillations. The results are depicted in \Fig{RegimeIIINonAd}. There, the oscillation frequency of the large spin is close to the Larmor frequency $\omega_{L} = B_z$, which appears in our depiction at a frequency $\tilde \omega_{L} = B_z/\lambda$. In the case without coupling between the two spin systems, the large spin oscillates exactly with the frequency $\tilde\omega_{L}$ in the $xy-$plane. The $z$-component is fixed due to $\ev{\dot J_z} = 0$ and possesses no coupling to the magnetic field.

The left graphs in \Fig{RegimeIIINonAd}, depict results for $\Gamma/\lambda = 10$. Here, the frequency of the electronic spin's $y$-component matches the frequency of the large spin's $x$- and $y$-component $ \omega_{S_y}=\omega_{J_x} = \omega_{J_y} \approx 0.8/\lambda $. But the  $z$-component of the electronic spin is almost twice the frequency of the other components, $\omega_{S_z} \approx 1.6\lambda$. This behavior survives for a smaller tunneling rate, see right graphs in \Fig{RegimeIIINonAd}, but there the frequency is even closer to the Larmor frequency, $\omega_{J_x,J_y,S_y } \approx 0.96\lambda$ and $\omega_{S_z} \approx 1.91\lambda$. The appearance of the frequency doubling for $\ev{\hat S_z(t)}$ is explained in \Sec{App.:FrequencyDoubling}.

\begin{figure}
  \begin{center}
    \includegraphics[width=0.4\textwidth]{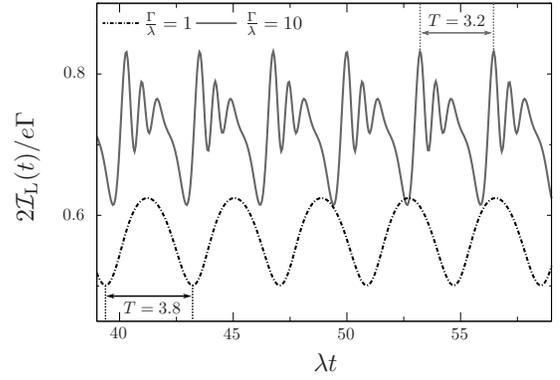}
  \end{center}
 \caption{\label{fig.:CurrentRegimeIII} Nonadiabatic current results corresponding to \Fig{RegimeIIINonAd}. The dashed-dotted line represents the current for a small tunneling rate, where smoothed oscillations appear. For higher $\Gamma$, solid gray line, the oscillation is still periodic but not longer sinusoidal. }
\end{figure}

The electronic current in regime III performs periodic oscillations. There, the results for the rate equation approach coincide with the nonadiabatic results, which are depicted in \Fig{fig.:CurrentRegimeIII} for the same parameters as in  \Fig{RegimeIIINonAd}.  
The frequency of the current oscillations, $\omega = 2\pi/T$, matches $\omega_{S_z}$. Not surprising, because this spin component corresponds to the occupation difference of the dot system, $2\ev{\hat S_z} = \ev{\hat n_{\uparrow}} - \ev{\hat n_{\downarrow}}$. Inserting the latter into the current equation \Eq{eq.:currentRate} and using the solution for the total occupation number in the long-time limit, cf. \Eq{eq.:SDStotaloccupationRate}, we obtain for the current through lead $\alpha$ in the rate equation frame
\begin{align}
 \mathcal I_{\alpha}(t) =&    e \left[  \left\{ \delta_{\alpha \rm L}
                              - \frac{\Gamma_{\rm L \uparrow} + \Gamma_{\rm L \downarrow}}{2 \Gamma}  \right\}
                                              (\Gamma_{\alpha \uparrow} + \Gamma_{\alpha \downarrow}) 
                              \right. \nonumber\\ & \left.
                              -   \ev{\hat S_z(t)} (\Gamma_{\alpha \uparrow} - \Gamma_{\alpha \downarrow}) \right]. 
\end{align}
And with $\Gamma = \Gamma_{\rm L \sigma} + \Gamma_{\rm R \sigma}$ it is obvious that $ \mathcal I_{\rm L}(t) = -  \mathcal I_{\rm R} (t)$ for all times $t$, and hence current conservation is ensured. The latter is also valid for the nonadiabatic approach.

From the above comparison, we can already conclude, that higher-order terms in the system-lead coupling, as included in the nonadiabatic approach, do matter in the regime of low magnetic field.

\section{Dynamics in the finite bias regime} \label{sec.:SDSfinitebias}
In the last section our investigation was focused on the infinite bias regime. There, the interaction between the large spin and the electronic system leads to interesting nonlinear effects. 
The rate equation approach is restricted to this regime of high external bias. With our nonadiabatic approach, we learned that differences appear if one includes higher-order transition terms. In this section, we take the next step by studying the finite bias regime. 

Due to the lack of an easy access to further analytic studies for the nonadiabatic approach, we use an adiabatic approach based on Keldysh Green's functions for the dynamical analysis; for details, see \Sec{App.:GreenFunctions}. This enables us to clarify the effects of a finite transport window. 

\subsection{Dynamical analysis : adiabatic approach} \label{Sec.Dynamicalanalyse}

Using a complete adiabatic approach for our system implies the assumption that the large spin's movement is slower compared to the electrons jumping through the system. This is an additional assumption on top of the mean-field approach, where quantum fluctuations are neglected already. But even for the derivation of the rate equations, one needs an adiabatic approximation for the electrons tunneling into the system. We performed the latter in the preceding section to decouple the lead-transition functions, cf. \Eq{eq.:SDSleadtransAdiabatic}. A full adiabatic approach also assumes that the electronic spin changes on a time-scale which is much smaller than that of the large spin. This approximation for the interaction between the electronic spin and the large spin reduces the number of dynamical equations to three, because only the ones for the large spin remain. The equations of motion for the electronic spin operators are solved with the help of Green's functions.

We can use the adiabatic approach to search for fixed points and also perform a rough characterization of them. The predictions for the classification of the fixed points coincide quite well with the actually obtained nonadiabatic results.
However, a complete adiabatic treatment of the system cannot capture the whole dynamics of the system. Even in the infinite bias regime, the adiabatic results do not coincide with those obtained from the rate equation or the nonadiabatic approach.

An improvement of this approach can be done by expanding the time-dependent Green's functions to first order, as done in Ref.\cite{Bode2012}for a related system. This expansion leads to additional friction terms in the equations of motions of the large spin. Then the dynamics are described by  a Landau-Lifshitz-Gilbert equation.\cite{Ralph2008}

\subsubsection{Fixed point analysis of $\mathcal P^{\pm}_{S0}$: center or saddle point}
The fixed points $\mathcal P^{\pm}_{S0}$, where the large spin is completely polarized parallel to the magnetic field, cf. \Eq{eq.:SDSfirstFIX}, appear also for the adiabatic system. 
But the corresponding value of the $z$-component for the electronic spin now
depends on several system parameters; for zero temperature it yields
\begin{align}\label{eq.:SzFirstFIX}
 \ev{\hat S_{z,0}(\pm j)} &= \sum_{\alpha} \left\{ \frac{\Gamma_{\alpha\uparrow}}{2\pi\Gamma} \arctan \left[ \frac{\mu_{\alpha} -\varepsilon_{\uparrow} (\pm j)}{\Gamma/2}\right] \right. \nonumber \\ & \left. \hspace{0.9cm}
                                        - \frac{\Gamma_{\alpha\downarrow}}{2\pi\Gamma} \arctan\left[ \frac{\mu_{\alpha} -\varepsilon_{\downarrow}(\pm j)}{\Gamma/2}\right]  \right\},
\end{align}
with $\varepsilon_{\uparrow,\downarrow}(j) = \varepsilon_d \pm 0.5 (B_z + \lambda j )$.
The $x$- and $y$-component are zero as for the rate equation approach. For infinite bias the $z$-component solely depends on the tunneling rates $\ev{\hat S_{z,0}^{\rm{IB}}} = (\Gamma_{\rm L \uparrow} -\Gamma_{\rm L \downarrow})/2 \Gamma$, which coincides with \Eq{eq.:SDSfirstFIX}.

For a further investigation of the fixed points, we use a linear stability analysis,\cite{Strogatz2000} where we study the Jacobian of the dynamical system around the fixed point. Some basic details of this analysis are denoted in Appendix \ref{App.:FixedPoints}, including the derivation of \Eq{eq.:SzFirstFIX}. 

For the fixed points $\mathcal P^{\pm}_{S0}$, one eigenvalue of the Jacobian (\ref{eq.AppJacobiFirstFIX}) is zero and the two others read
\begin{align} \label{eq.:SDSeigenvaluesFIX1adiabatic}
 \mathcal E_{2,3}^{0,\pm} =& \pm i \sqrt{ \mathcal T_0
                                  \left( \mathcal T_0
                                 \mp \lambda j \left. \frac{\partial  \ev{\hat S_x}}{\partial  \ev{\hat J_x}} \right|_{\substack{ \ev{\hat J_{x,0}}=0,\\ \ev{\hat J_{z,0}}=\pm j} }  \right) }, \nonumber
                                  \\ \mbox{with} & \ \
  \mathcal T_0 =    \left[ \lambda  \ev{\hat S_{z,0}(\pm j)} + B_z \right].
                                 \end{align}
Therewith, two possible realizations can appear: either the eigenvalues are purely imaginary, classifying the fixed point as a stable center, or purely real, corresponding to a saddle point. In the stable center case, the spin trajectories perform periodic oscillations around the fixed point and their amplitudes are determined by the initial conditions. In contrast, no oscillations appear if the fixed point can be classified as a saddle point, where the trajectories get repelled and approach other stable solutions of the dynamical system.

From \Eq{eq.:SDSeigenvaluesFIX1adiabatic} we see, that no damped oscillations around the fixed points $\mathcal P^{\pm}_{S0}$ are expected in the adiabatic regime, because this so-called stable spiral case would require complex eigenvalues with a finite negative real part.
This missing damping is characteristic for the adiabatic results, hence this approach cannot describe the complete dynamics of the system. The latter is valid even in the infinite bias regime, since the appearing dynamical features such as self-sustained oscillations require both,  positive and negative damping, \cite{Hussein2010} whereby positive damping is indicated by a positive real part of the eigenvalue.

\begin{figure}[t]
	\centering
		 \includegraphics*[width=1.0\linewidth]{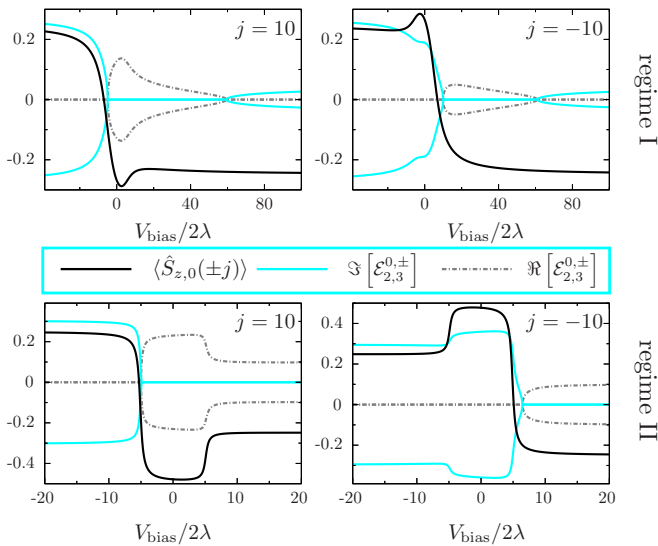}
  	 \caption{Results of the dynamical analysis for region I and II. Here, parameters are chosen as in \Sec{Sec.:BerlinMadrid}. The solid black line denotes the $z$-component of the electronic spin $\ev{\hat S_z}$ and the external bias is chosen in a symmetric manner $\mu_{\rm L} =-\mu_{\rm R}= V_{\rm bias}/2$. 
  	 Zero imaginary part appears in region I for alignment of $\ev{\mathbf{\hat J}}$ in the direction of $B_z$ in the range of $V_{\rm bias}/2\lambda \in [-5; 60]$ and for alignment in the opposite direction the range yield $V_{\rm bias}/2\lambda\in [10; 60]$. For region II the imaginary part disappears in the range of $V_{\rm bias}/2\lambda \in [-5; \infty]$ for  $\ev{\hat J_{z,0}} = j $  and $V_{\rm bias}/2\lambda \in [6.4; \infty]$ for $\ev{\hat J_{z,0}} = -j $.}
	   \label{FIX1bias}
\end{figure}

In \Fig{FIX1bias}, $\ev{\hat S_{z,0}}$ as a function of the applied bias is depicted, together with the real and the imaginary part of the corresponding eigenvalue. The two upper graphs show the behavior in regime I for $\ev{\mathbf{\hat J}}$ pointing in the direction of the magnetic field (left, $j=10$) as well as against it (right, $j=-10$).
We observe regions where the imaginary part of the eigenvalues is equal to zero and therewith the oscillations disappear. 

The ranges of the finite real part are slightly different for $\mathcal P^{+}_{0}$ and $\mathcal P^{-}_{0}$. These differences appear for small bias; the intervals of the finite real part are denoted in the caption of \Fig{FIX1bias}. For $\mathcal P^{-}_{S0}$, the intervals with a finite real part are smaller than those for $\mathcal P^{+}_{S0}$. Depending on the existence of other fixed points, the regions where only one fixed point has eigenvalues with finite imaginary parts, are promising for the occurrence of spin-flips.

Based on the chosen polarization of the leads and therewith the different tunneling rates for the spin-down electrons, we obtain a system which is not symmetric. Hence, the dynamical behavior for negative detuning is different from that for positive detuning, $V_{\rm bias}>0$. This is clearly visible in \Fig{FIX1bias}, where for large negative bias the imaginary part in region I is always unequal to zero. The infinite bias result for region I equals $\ev{\hat S_{z,0}}^{IB} = \pm 0.25 $ as expected, whereby the sign depends on the detuning. For region I, the system performs periodic oscillations for a bias $V_{\rm bias}/2\lambda > 60$.

In contrast, in region II (2nd row in \Fig{FIX1bias}), no finite imaginary part for the high bias regime exists. This coincides with the infinite bias result for the nonadiabatic approach, where the concerned fixed points do not appear as stable centers or spirals. But here, for a small bias range and also for negative detuning, we find, that the eigenvalues can become complex and therewith the fixed points $\mathcal P^{\pm}_{S0}$ can exist as centers. For positive detuning, the fixed point $\mathcal P^{+}_{S0}$has already turned into a saddle point, but the $\ev{\hat J_{z,0}} = -10$ state is alive until $V_{\rm bias}/2\lambda = 6.4$. 

\begin{figure}[t]
	\centering
		 \includegraphics*[width=1.0\linewidth]{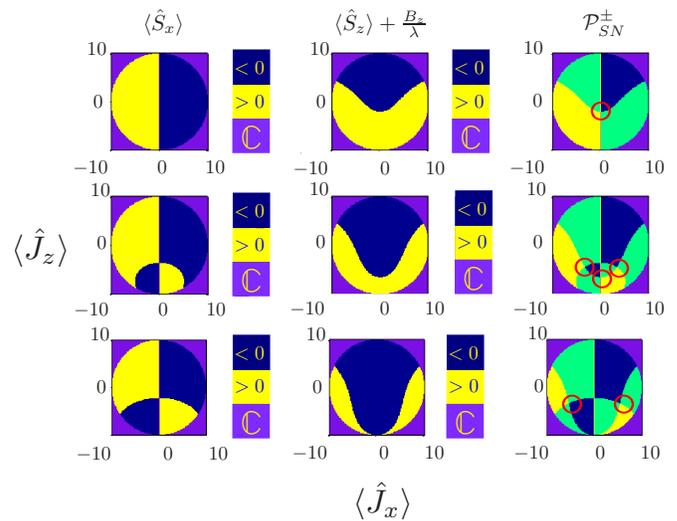}
  	 \caption{Results for the fixed points $\mathcal P^{\pm}_{SN}$ in region I: numerically calculated density plots for $\ev{ \hat S_z} + B_z/\lambda$ and $\ev{ \hat S_x}$ as a function of the large spin components. These functions must be both equal to zero 
  	 for the appearance of $\mathcal P_{\pm}^{SN}$. 
  	 The yellow (blue) area in the first two columns corresponds to positive (negative) values. Along the borders of these areas the functions are equal to zero.
  	 The right column depicts the superposition of the first two columns, there a contact of the blue and yellow area denotes a solution for $\mathcal P^{\pm}_{SN}$ (red circles). The bias increases from the upper to the lower row, explicit values are $V_{\rm bias}/2\lambda = 5;8;10$. 
  	 The numerical results were sorted by their sign, with the purpose of highlighting the positive and the negative areas. 
  	 The bifurcation appears approximately at $V_{\rm bias}/2\lambda \approx 7$.}
	   \label{SPNplotRegimeI}
\end{figure}

For region III, which means for $B_z/\lambda > 0.25$, the behavior of the fixed points is similar to region I, if the magnetic field and the tunneling rate are small. But for higher values of these parameters, the real part of the eigenvalues is always zero and therewith the fixed points can be classified as stable centers with purely imaginary eigenvalues. The transition from purely real to purely imaginary eigenvalues is only numerical accessible. In  \Fig{SDSFixDensity}  we depict the appearance of stable solutions as a function of the magnetic field and the tunneling rate for different bias values. In these graphs, the discussed transition is visible.

\subsubsection{Two conditions for the fixed points $\mathcal P^{\pm}_{SN}$}

The next fixed points obtained from \Eq{eq.:LargeSpinEOM} include the requirement that  $\ev{\hat S_x} = 0$ and the condition $\ev{\hat S_z} + B_z/\lambda = 0$ has to  be complied with as well. We define
 \begin{align}
 \mathcal P^{\pm}_{SN}: \left( \ev{\hat J_{x,N}},\pm\sqrt{j^2- \ev{\hat J_{x,N}}^2-  \ev{\hat J_{z,N}}^2}, \ev{\hat J_{z,N}}\right).
 \end{align}

In the infinite bias regime $\mathcal P_{SN}^{\pm}$ coincide with the four fixed points $\mathcal P_{S1,S2}^{\pm}$ in region II, see \Eq{eq.:SDSsecondFIX} and below, obtained from the rate equations. 
\begin{figure}[t]
	\centering
		 \includegraphics*[width=1.0\linewidth]{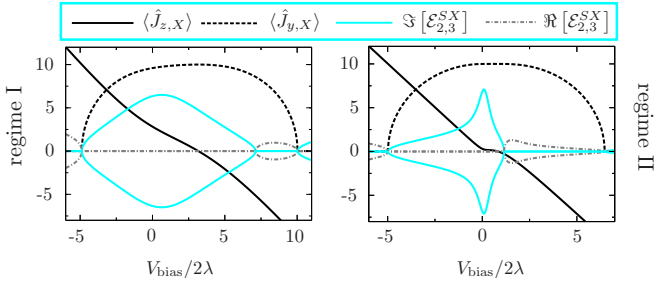}
  	 \caption{Results of the dynamical analysis of $\mathcal P^{+}_{SX}$ for region I and II. Outside of the depicted range, the fixed point is complex. The real range is limited by $\ev{\hat J_{y,X}}$, which is only real inside of the dashed line. Parameter as in \Sec{Sec.:BerlinMadrid}.}
	   \label{FIX2bias}
\end{figure}
The investigation of $\mathcal P^{\pm}_{SN}$ in the finite bias regime is possible only numerically. In \Fig{SPNplotRegimeI}, we present results obtained for region I. The cycle shapes originate from the conservation of the large spin, hence the radius equals $j$. 
Outside of these cycles, $\ev{\hat J_y}$ becomes complex and no real solution for $\mathcal P^{\pm}_{SN}$ exists.

The stability is defined by the eigenvalues of the Jacobian (\ref{eq.JacobianFull}) evaluated at $\mathcal P^{\pm}_{SN}$. One eigenvalue of  $\mathcal P_{SN}^{\pm}$ is again zero, due to the fact, that we actually deal with a two-dimensional system. The remaining eigenvalues yield
\begin{align} \label{eq.:Eigenvaluesgeneral}
\mathcal E_{2,3}^{SN} =&  \frac{\lambda}{2} \ev{\hat J_y}  \left\{
\left( \frac{\partial  \ev{\hat S_x}}{\partial  \ev{\hat J_z}} - \frac{\partial  \ev{\hat S_z}}{\partial  \ev{\hat J_x}}\right) 
 \right. \nonumber \\ & \pm \left.
\sqrt{ \left(\frac{\partial  \ev{\hat S_x}}{\partial  \ev{\hat J_z}} + \frac{\partial  \ev{\hat S_z}}{\partial  \ev{\hat J_x}}\right)^2 
    - 4  \frac{\partial  \ev{\hat S_x}}{\partial  \ev{\hat J_x}}  \frac{\partial  \ev{\hat S_z}}{\partial  \ev{\hat J_z}} }
\right\} \Bigg|_{\mathcal P^{\pm}_{SN}}.
\end{align}

For a small bias, $V_{\rm bias}/2\lambda < 5$, only one solution for $\mathcal P^{\pm}_{SN}$ exists, corresponding to the case $\ev{\hat J_{x,N}} = 0$, which we name $\mathcal P^{\pm}_{SX}$. The latter is stable until $V_{\rm bias}/2\lambda < 7$ cf. \Fig{FIX2bias}.
By further increasing the bias another bifurcation appears, where $\mathcal P^{\pm}_{SX}$ becomes unstable and two other fixed points are created. These points are symmetric to the $\ev{\hat J_x}$ - axis and move with higher bias values further to $\ev{\hat J_z} = \pm j$. When they reach the border of the cycle, they disappear and \mbox{$\ev{ \hat S_z} + Bz/\lambda = 0$} no longer has a real solution. The emerging fixed points have negative/positive real eigenvalues and can be characterized as stable/unstable nodes. For region II, these points do not disappear for high bias and the nodes are the only remaining stable solutions of the system. These nodes correspond to fast damping behavior for the spin components.

For $\mathcal P^{\pm}_{SX}$, where $\ev{\hat J_{x,N}} \equiv \ev{\hat J_{x,X}} =0$, the last term in \Eq{eq.:Eigenvaluesgeneral} is unequal to zero and the component $\ev{\hat J_{z,X}}$ is obtained from the transcendental equation
\begin{align}\label{eq.:FIXjxNullconsition}
 - \frac{B_z}{\lambda} =&  \sum_{\alpha} \left\{ \frac{\Gamma_{\alpha\uparrow}}{2\pi\Gamma} \arctan 
                          \left[ \frac{\mu_{\alpha} - \varepsilon_d - \frac{1}{2} (B_z + \lambda \ev{\hat J_{z,X}} )}
                                      {\Gamma/2}\right] 
                           \right. \nonumber \\ & \left. 
                               - \frac{\Gamma_{\alpha\downarrow}}{2\pi\Gamma} \arctan
                          \left[ \frac{\mu_{\alpha}  - \varepsilon_d + \frac{1}{2} (B_z + \lambda \ev{\hat J_{z,X}} )}{\Gamma/2}\right] \right\},
\end{align}
This equation has no solution for $\ev{\hat J_{z,N}}$ in the infinite bias case and therefore the fixed points $\mathcal P^{\pm}_{SX}$ do not exist there. In the regime of a low magnetic field, the right side of \Eq{eq.:FIXjxNullconsition} has to be small. Following from that, the argument of the $\arctan$ function has to be  small as well, leading to the estimate for the evolution of the $z$-component as linear to the applied bias, $\ev{\hat J_{z,X}} \sim V_{\rm bias} $.

This linear behavior is clearly visible in \Fig{FIX2bias}, where we plotted $\mathcal P^{+}_{SX}$ and its eigenvalues as a function of the applied bias. The arcs correspond to the $\ev{\hat J_{y,X}}$ components, which determines whether the fixed point exists 
or not. Within these arcs, the point is real and therewith physically reasonable. The radius of the arcs are limited by the length of the large spin. 

For both regions we observe small ranges with a finite imaginary part, where the fixed point can be classified as a stable center. We also observe ranges where a saddle point occurs. In regime I, the point $\mathcal P^{+}_{SX}$ starts its existence for $V_{\rm bias}/2\lambda \approx -5$. At this bias value, $\mathcal P^{+}_{S0}$ turns into a saddle point, cf. \Fig{FIX1bias}. 

This behavior agrees quite well with that of $\mathcal P^{\pm}_{S0}$. We can interpret $\mathcal P^{\pm}_{SN}$ as the complementary points in the region, where $\mathcal P^{\pm}_{S0}$ disappear. As we discussed before, for a certain bias region the stable solutions $\mathcal P^{\pm}_{S0}$ turn into saddle points, and we can estimate, that then the fixed points appearing in \Fig{SPNplotRegimeI}, correspond to stable solutions of the dynamical system.

\subsubsection{Fixed points for unpolarized leads}
\begin{figure}[t]
	\centering
		 \includegraphics*[width=1.0\linewidth]{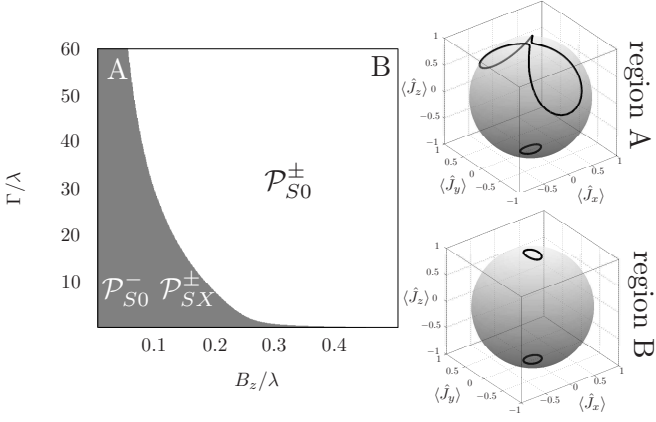}
  	 \caption{Fixed points for unpolarized leads as a function of tunneling rate $\Gamma$ and magnetic field $B_z$. The symmetric bias is chosen as $V_{\rm bias}/2\lambda = 5$. 
  	  The right graphs depict the adiabatic results for the large spin, in region B the small cycles on the top/bottom correspond to $\mathcal P^{\pm}_{S0}$. In region A, instead of  $\mathcal P^{+}_{S0}$ the fixed point $\mathcal P^{+}_{SX}$ appears (larger cycle). Magnetic field for region A/B is $B_z/\lambda = 0.1/0.3$.}
	   \label{UnpilBias10plot}
\end{figure}
For unpolarized leads, the tunneling rates become spin-independent, $\Gamma = \Gamma_{\rm L} + \Gamma_{\rm R}$. 
There only the fixed points $\mathcal P^{\pm}_{S0}$ and $P^{\pm}_{SX}$ exist, as depicted in \Fig{UnpilBias10plot}. In dependence on the tunneling rate and the magnetic field, two main regions appear. The first one, region A, has the stable fixed points $\mathcal P^{-}_{S0}$ and $P^{\pm}_{SX}$, which can be characterized as centers. In this region, spin-flips of the large spin can be possible, because  $\mathcal P^{+}_{S0}$ is no stable solution. But the system is still multistable, and therewith the trajectories can also end up in the fixed points $P^{\pm}_{SX}$. Note, that for a total adiabatic ansatz, a complete spin-flip is not observable, because the trajectories starting parallel to the magnetic field end up in $P^{\pm}_{SX}$, as shown in the right graphs of \Fig{UnpilBias10plot}. If one reverses the direction of the magnetic field, region A contains $\mathcal P^{+}_{S0}$ instead of $\mathcal P^{-}_{S0}$.

In region B, only $\mathcal P^{\pm}_{S0}$ are stable fixed points. By further increasing the bias, the region where $P^{\pm}_{SX}$ exist gets smaller, and in the infinite bias case only the second region persists and all electronic spin components $\ev{\hat S_i}$ become zero.
Then the two systems decouple and the large spin oscillates with the Larmor frequency.

\subsection{Results for the nonadiabatic approach in the finite bias regime} \label{Sec:SDSfiniteBiasresults} 
We expect to obtain the same fixed points for the nonadiabatic approach as for the adiabatic approach. For the case $\mathcal P^{\pm}_{S0}$, we want to show, that these fixed points also appear in the nonadiabatic regime. There, all lead-transition functions for different spin $\mathcal T_{\sigma\sigma'}^{\alpha}$ vanish in the stationary case and we obtain for the ones with equal spins the simple result
\begin{align}
 \mathcal T_{\sigma\sigma}^{\alpha}(\omega) =  i \frac{\Gamma_{\alpha \sigma}}{2\pi} \frac{f_{\alpha}(\omega)}{(\omega -\varepsilon_{\sigma}(\pm j) + i\frac{\Gamma}{2})} = i \frac{\Gamma_{\alpha \sigma}}{2\pi} G^{R}_{\sigma}(\omega),
\end{align}
containing the spin-dependent single-level Green's function $G^{R}_{\sigma}(\omega)$ without coupling of the two electronic levels, due to $\ev{\hat J_{x,0}}=0$. Also, the equation for the $z$-component of the electronic spin is straightforwardly obtained from
\begin{align}
 \ev{\hat S_{z,0}} =& \frac{1}{\Gamma} \sum_{\alpha} \int d\omega \Re\left[ \mathcal T^{\alpha}_{\uparrow \uparrow}(\omega) - \mathcal T^{\alpha}_{\downarrow \downarrow}(\omega) \right] \nonumber \\ 
 =& \frac{1}{2\pi} \sum_{\sigma \alpha} \frac{\Gamma_{\alpha \sigma}}{\Gamma} \left[2\delta_{\sigma\uparrow} - 1 \right]
    \arctan \left[\frac{\mu_{\alpha} - \varepsilon_{\sigma}(\pm j) }{\Gamma/2} \right].
\end{align}
This result coincides with \Eq{eq.:SzFirstFIX}, and the fixed points are identical to the ones obtained with the adiabatic approach. In the same manner, we can construct also the other fixed points.  In the general stationary case, all lead-transition functions decouple from the equation system and can be expressed by
the retarded Green's function. The adiabatic expressions for electronic spin components can also be reconstructed.   

This equivalence is not valid considering the adiabatic eigenvalues; they are not directly transferable to our nonadiabatic approach. The fixed points are long-time quantities and stationary solutions of the dynamical system, in contrast the eigenvalues include higher-order terms, and their calculation requires the first derivations of the system's variables.    

But some predictions derived from the adiabatic eigenvalues also appear in the nonadiabatic approach. Therefore, we can estimate that the behavior for the full time-dependent solution is strongly influenced by the adiabatic eigenvalues. 

\subsubsection{Region I: Disappearance of the oscillations}

We start our presentation and discussion of the nonadiabatic results with the ones obtained in region I, corresponding to a low value of the external magnetic field and a large tunneling rate. In the case of infinite bias, we obtained an oscillating behavior between two cycles.
This is different from the rate equation results, where damped oscillations appear. 

From the adiabatic analysis of the preceding section, 
we obtained the prediction, that the fixed point $\mathcal P^{+}_{S0}$ of the system turns from a center into a saddle point for the bias range $V_{\rm bias}/2\lambda \in [-5;60]$. When $\mathcal P^{+}_{S0}$ loses its stability at $V_{\rm bias}/2\lambda \approx -5$, a supercritical pitchfork bifurcation \cite{Kuznetsov,Beuter2003,Strogatz2000} appears and $\mathcal P^{\pm}_{SX}$ are created; they exist until $V_{\rm bias}/2\lambda \approx 7$ as centers. If the latter fixed points become unstable, two stable nodes are born and we expect the oscillations to disappear. Note, that these predictions  originate from the adiabatic eigenvalues. 

\begin{figure}[t]
	\centering
		 \includegraphics*[width=1.0\linewidth]{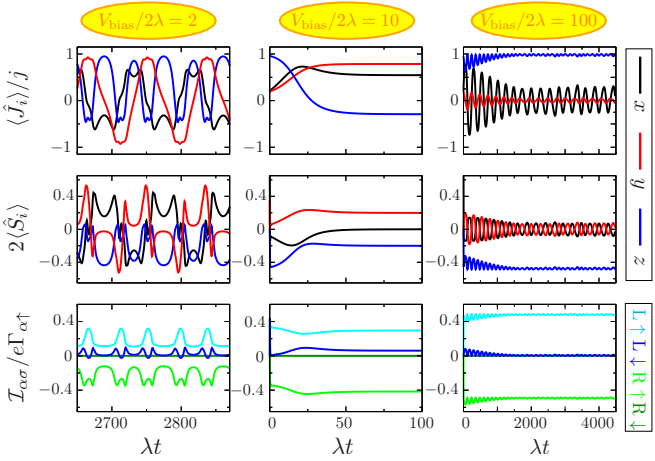}
  	 \caption{Nonadiabatic results obtained in region I for three different bias values, which are chosen in a symmetric manner $V_{\rm bias} = \mu_{\rm L}/2= -\mu_{\rm R}/2$. The electronic current, depicted in the lowest row, is separated into left/right and spin-up/-down contributions. The magnetic field yields $B_z/\lambda=0.1$ and $\Gamma/\lambda = 9$. Initial conditions as in  \Sec{Sec.:BerlinMadrid} for regime I.}
	   \label{FBRegimeI.eps}
\end{figure}

As depicted in \Fig{FBRegimeI.eps}, we obtain these features also in our nonadiabatic results. There, the spin components and the current are depicted for three different bias values. For $V_{\rm bias}/2\lambda = 2 $, we observe non-sinusoidal self-sustained oscillations.  
$\ev{\hat S_z}$ oscillates around $-B_z/\lambda$, which corresponds to a stable point of kind $\mathcal P^{\pm}_{SX}$.

If we increase the applied bias for the initial conditions corresponding to the results depicted in \Fig{FBRegimeI.eps}, the oscillations disappear in the range of $V_{\rm bias}/2\lambda \approx 8$. They run into the fixed point of the kind $\mathcal P^{+}_{SN}$, which is clearly visible in \Fig{FBRegimeI.eps}, where $\ev{\hat S_x}=0$ and $\ev{\hat S_z}=- B_z/\lambda$. 

By varying the initial conditions, we observe that not all oscillations have disappeared for a bias in the range of $V_{\rm bias}/2\lambda \approx 8$. Choosing the initial conditions near the fixed point $\mathcal P^{-}_{S0}$, we observe oscillations,  which are smoothly sinusoidal and run around $\mathcal P^{-}_{S0}$. For $V_{\rm bias}/2\lambda \approx 10$, the oscillations disappear and the trajectories run in the same fixed point as depicted in the middle graph of \Fig{FBRegimeI.eps}. This result coincides perfectly with the border predicted in the adiabatic analysis.

The bifurcation point, where a revival of the oscillations appears, is also correctly predicted by the adiabatic analysis. For $V_{\rm bias}/2\lambda \approx 60$, we observe again oscillations around $\mathcal P^{\pm}_{S0}$. In these oscillations, there is a slight indication of the two cycle behavior, as in the infinite bias case, but the difference between the radii of the cycles is not large. The latter is visible in the third column of \Fig{FBRegimeI.eps}, where the results for  $V_{\rm bias}/2\lambda = 100 $ are depicted.

The lower row in \Fig{FBRegimeI.eps} depicts the electronic current $\mathcal I_{\alpha\sigma}(t) $, separated into its constituent parts. The right tunneling amplitude is equal to zero, and following from that is the corresponding current channel. Therefore, the current $\mathcal I_{\rm R \uparrow}$ should be equal to the total current through the system, hence current conservation remains valid.
For the nonadiabatic current, this is ensured for the time-averaged current values, but it can be different in the time-dependent case. There, we observed some accumulation of current in the central region. 

When we consider the current channels for $V_{\rm bias}/2\lambda = 2$ in \Fig{FBRegimeI.eps}, we notice that all channels reach their minima if the $\ev{\hat J_z}$-component is maximal and therewith here close to $j$. For this low bias regime, the shift of the energy level, due to the coupling to the large spin $ \pm 1/2 (B_z \lambda + \ev{\hat J_z}) $, leads to a positioning of the levels slightly outside of the transport window. Hence the current is minimal.

\begin{figure}[t]
	\centering
		 \includegraphics*[width=1.0\linewidth]{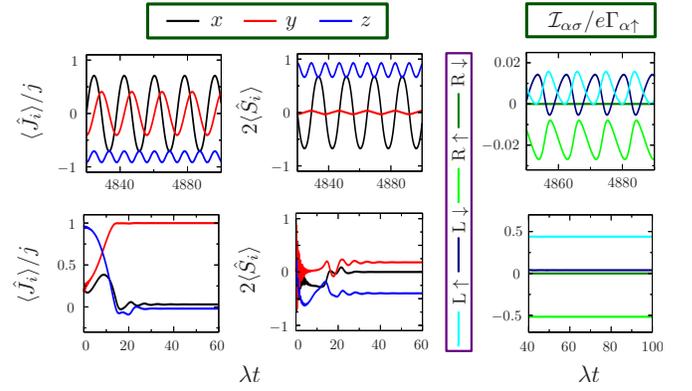}
  	 \caption{Nonadiabatic results for region II. The upper row corresponds to a small bias $V_{\rm bias}/2\lambda = 2$, where the spin components oscillate around the coordinates of $\mathcal P^{-}_{S0}$. Increasing the bias, leads to a disappearing of the oscillations and one fixed point $\mathcal P^{+}_{SN}$ is present, see lower row for $V_{\rm bias}/2\lambda = 5$. The magnetic field yields $B_z/\lambda=0.2$ and $\Gamma/\lambda = 0.7$. Note, that the time-intervals of the current and the spin operators differ slightly.
         Initial conditions as in  \Sec{Sec.:BerlinMadrid} for regime I.}
         \label{FBregimeIIa}
\end{figure}

The current corresponding to the spin-up electrons flows in the direction of the bias. For the spin-down electrons, the situation is more complicated, because they are not allowed to leave the system through the right lead due to $\Gamma_{\rm{R}\downarrow}=0$. The electrons can either flip their spin, stay in the lower level, or flow back into the left lead. If the last process appears the electron moves against the bias and the current becomes negative. This feature is slightly visible in the lowest graph of the first column in \Fig{FBRegimeI.eps}, where the current $\mathcal I_{\rm L \downarrow}$ drops below zero for a quite small region.

Again, we can address this effect on the position of the large spin's $z$-component. In the case $\mathcal I_{\rm L \downarrow}<0$, the shifting of the energy level is the other way around, because $\ev{\hat J_z}$ is negative.  Following from that, the spin-down level lies above the spin-up level and additionally in the neighborhood of the left Fermi edge ($\mu_{\rm L} = 2/\lambda$), and electrons can occupy empty states in the left lead. After $\ev{\hat J_z}$  passes its minima, the spin-down level moves down and therewith its current channel drops as well. 

\subsubsection{Region II: Negative Detuning and spin-flip of the large spin}

The feature of negative current is more visible in region II, as we see in the current results for $V_{\rm bias}/2\lambda = 2$, which are depicted in the upper row of \Fig{FBregimeIIa}. There, the large spin's $z$-component oscillates close to $-j$ and therewith, both effective levels are clearly situated outside the transport window and the spin-down level lies again above the spin-up level. The oscillations of $\ev{\hat J_z}$ are comparatively small and do not influence the behavior of the effective levels as much as for region I. They stay outside of the transport window for all times.

We try to interpret the evolution in time for the current channels focusing on the transition between the levels. If $\mathcal I_{\rm L \downarrow }$ is negative, the right current reaches its minimum and the left current for spin-up electrons is maximal. Due to $\mathcal I_{\rm R \uparrow } < \mathcal I_{\rm L \uparrow} $, we can assume that spin-flips from the lower (spin-down) to the upper level (spin-up) happen, and a depletion of the upper level into the left lead appears, due to the negativity of $\mathcal I_{\rm L\downarrow}$. This is in accordance with the evolution for $\ev{\hat S_z}$, which decreases in this range.

In contrast, when $\ev{\hat S_z}$ increases, the left current for spin-down electrons is maximal, as well as $\mathcal I_{\rm R \uparrow }$. But $\mathcal I_{\rm L\uparrow }$ is close to zero in this regime. Therefore, we interpret this kind of current cycle in the following way. A spin-down electron enters the upper level, flips into the lower level, due to the interaction with the large spin, and finally leaves the central system to the right lead. This would explain why the current for spin-up electrons is maximal at the right lead and minimal at the left lead.  

\begin{figure}[t]
	\centering
		 \includegraphics*[width=1.0\linewidth]{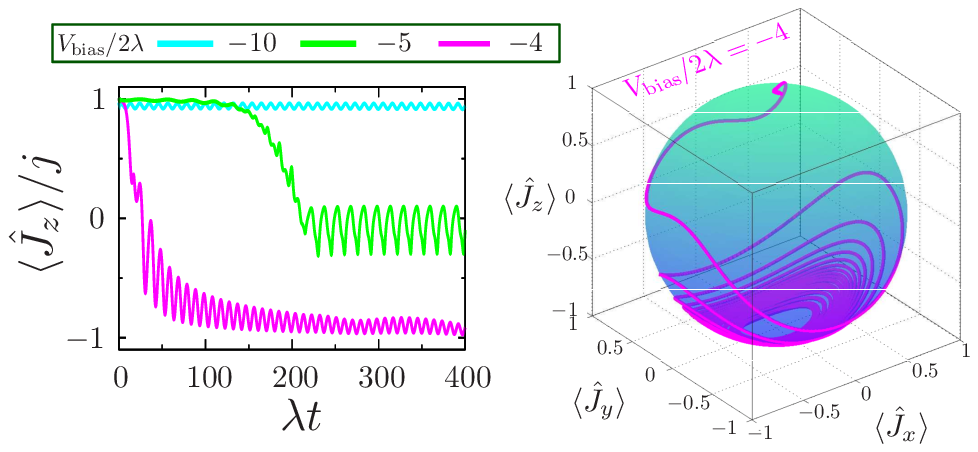}
  	 \caption{The large spin's $z$-component for several bias values in regime II. Parameters as for \Fig{FBregimeIIa}, and initial conditions are $\ev{\hat J_x}= \ev{\hat J_y} = \sqrt{50-9.9^2/2}, \ev{\hat J_z}= 9.9$ and $\ev{\hat S_y}= \ev{\hat S_z} = 0, \ev{\hat S_x}= 0.5$. }
	   \label{DetuningRegimeIIa}
\end{figure}

If the bias is increased, the regions of negative spin-down current vanish, as do the oscillations of the spin components, as depicted in the lower row of \Fig{FBregimeIIa}. Remember, for the infinite bias regime, we found no periodic oscillations at all in this region. In accordance with the foregoing adiabatic analysis, the  trajectories enter one fixed point of the kind $\mathcal P^{\pm}_{SN}$.

The adiabatic analysis, also predicts an earlier disappearance of the fixed point $\mathcal P^{+}_{S0}$ than the fixed point $\mathcal P^{-}_{S0}$. As a result, we propose the possibility of spin-flips for the large spin in the region, where only the eigenvalues of $\mathcal P^{-}_{S0}$ have a finite imaginary part: $V_{\rm bias}/2\lambda \in [-5; 6.4]$. 

To monitor how the first fixed point disappears, we choose our initial conditions close to $\mathcal P^{+}_{S0}$ and assume negative detuning. The results are depicted in  \Fig{DetuningRegimeIIa}. For $V_{\rm bias}/2\lambda=-10$, the trajectories perform smooth oscillations around $\mathcal P^{+}_{S0}$. As expected from \Fig{FIX1bias}, the $z$-component of the electronic spin $\ev{\hat S_z} \approx 0.2$ and is therewith positive. As discussed in the last section, the observed system is not symmetric and hence the fixed point $\mathcal P^{+}_{S0}$ stays alive by further decreasing the bias.

Increasing the bias, we see a different behavior, i.e., $\ev{\hat J_z}$ drops down when the bias passes the threshold $V_{\rm bias}/2\lambda=-5$, as is clearly visible in \Fig{DetuningRegimeIIa}. At first the trajectory runs into a fixed point of the kind $\mathcal P^{\pm}_{SX}$, laying in the middle of both points $\mathcal P^{\pm}_{S0}$. But even for $V_{\rm bias}/2\lambda=-4$, the spin components enter the fixed point $\mathcal P^{-}_{S0}$. 

\begin{figure}[t]
	\centering
		 \includegraphics*[width=1.0\linewidth]{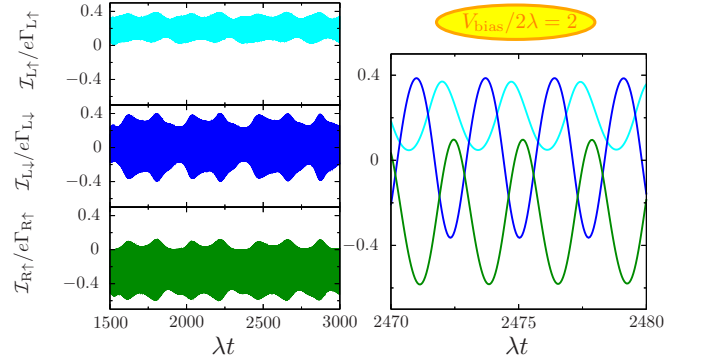}
  	 \caption{The graphs depict the different contributions to the current in regime III. The right graph presents a detail of the left graphs drawn together for a short time range. Initial conditions and parameters as in  \Sec{Sec.:BerlinMadrid} for regime III.}
	   \label{FBRegimeIIIb}
\end{figure}

\begin{figure*}[t]
	\centering
		 \includegraphics*[width=\linewidth,clip]{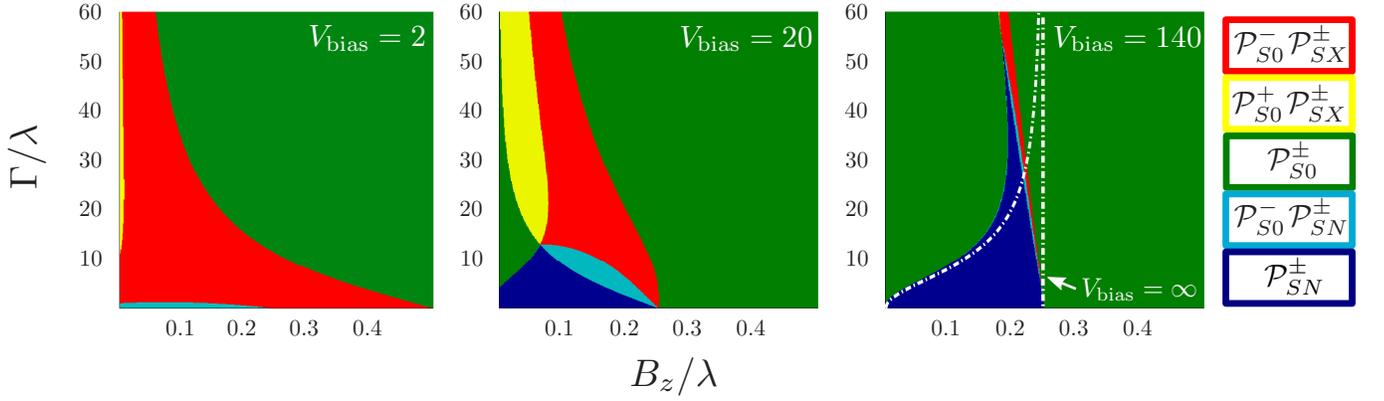}
  	 \caption{Stable fixed points as a function of tunneling rate and magnetic field. The external bias is increased from left to right, the explicit values are denoted in the graphs (in units of $[\lambda]$). Additionally, the infinite bias result is plotted in the last graph on the right side. The dark blue area denotes the regime where no oscillations appear and only the nodes $\mathcal P^{\pm}_{SN}$ exist.
  	  }
	   \label{SDSFixDensity}
\end{figure*}

If we decrease the tunneling rate, we observe a transition to chaotic-like oscillations as in the infinite bias case; see \Fig{RegimeIINonAd}.
For small values of the bias, the chaotic-like behavior is a little suppressed and the trajectories oscillate comparatively smoothly with a high frequency.

\subsubsection{Region III: Oscillations and high frequency}

In \Sec{Sec.:BerlinMadrid} we presented results for larger values of the magnetic field, where the trajectories oscillate around $\mathcal P^{\pm}_{S0}$. This kind of behavior is recovered for the finite bias regime. But we observe slight differences in the regime of small bias.
In \Fig{FBRegimeIIIb}, the current for regime III is depicted ($V_{\rm bias}/2\lambda =2$); the results for the left and the right current are split into their contributions from spin-up and -down electrons. 
We observe that the current channel $\mathcal I_{\rm{L} \downarrow}$ oscillates around the zero axis and, following from that, the current is flowing in both directions. The frequency of the oscillations is quite high, and the spin components oscillate between two cycles, whose signatures are clearly visible in the left graphs of \Fig{FBRegimeIIIb}.

\section{Conclusion} \label{Sec.Conclusion}

\begin{figure}[t]
	\centering
		 \includegraphics*[width=\linewidth,clip]{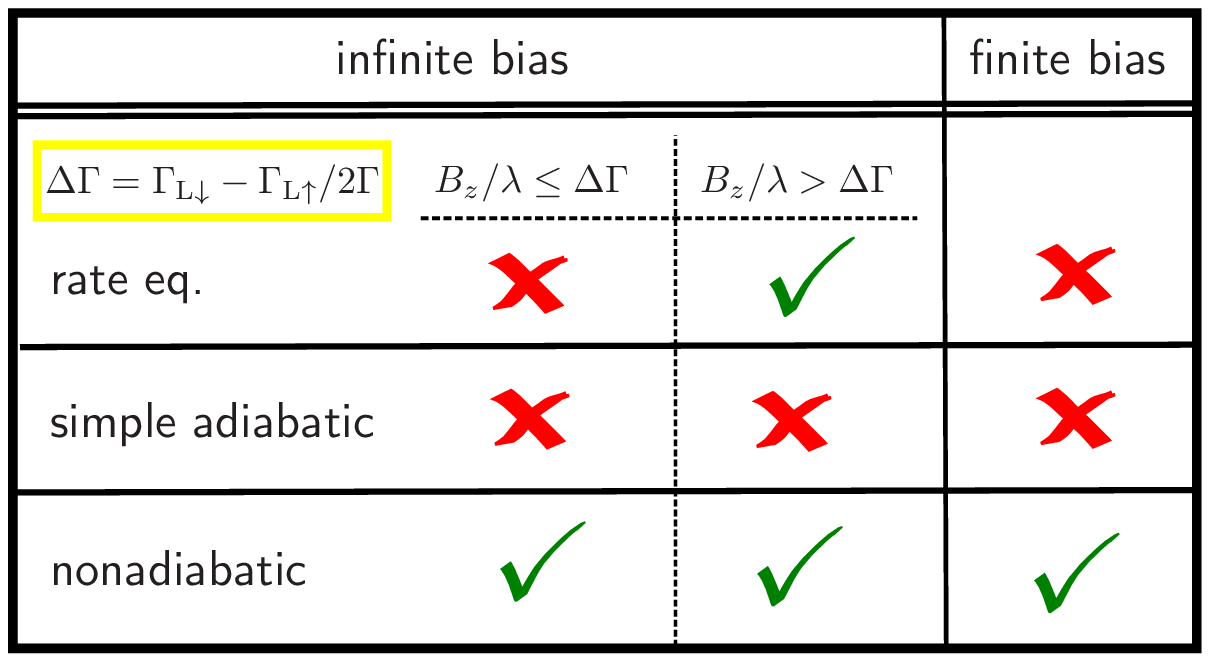}
  	 \caption{Table listing the regions were the different approaches are valid (green check mark) or not valid (red cross). 
  	 $\Delta \Gamma $ corresponds to the border between region II and region III, i.e. in our foregoing calculations we had $\Delta \Gamma = 1/4$.
  	 Note, that this table corresponds to a conservative estimate focusing on parameter regimes where the approaches always coincide with the nonadiabatic approach and for polarized leads.
  	  }
	   \label{Fig.Tabelle}
\end{figure}

With our nonadiabatic approach, we are able to probe the rate equation and the adiabatic approach and thus can discuss the advantages and disadvantages of these two methods, cf.\Fig{Fig.Tabelle}.  

The rate equation approach from \cite{Lopez-Monis2012}, is a quite practical method. The numerical effort is low and the dynamical system can be investigated analytically. However, this method neglects higher-order transitions, and we learned from the nonadiabatic results, that these are relevant in the regime of low magnetic field. In the latter region, the rate equations miss parts of the nonlinear dynamics. If the magnetic field increases, the results of both approaches coincide.

The limitation of the rate equation approach to high external bias is another disadvantage: at finite bias, we observe even richer dynamics within the nonadiabatic approach. There, features such as spin-flips of the large spin and suppression of the oscillations as a function of the applied bias appear. The disadvantages belonging to the nonadiabatic approach are the high numerical effort and its inaccessibility for further analytic investigations. 

In our work, we utilized the adiabatic Green's function method to analyze our system. However, we omitted the presentation of time-dependent results, because this simple adiabatic approach does not capture a lot of the dynamics even in the infinite bias regime. 
There, the dynamical system reduces to three equations of motion, and only if the trajectories run directly in a stable node and no oscillations appear can the adiabatic approach correctly describe the system. 

On the other hand, the adiabatic Green's functions are suitable for the analysis of the system. For the finite bias regime, our initial adiabatic analysis matched quite well the behavior of the nonadiabatic approach. Therewith, we can re-define the dynamical regions as depicted in \Fig{SDSFixDensity}. It is clearly visible that the external bias is an important parameter for this system. The regions where only one of the main fixed points $\mathcal P^{\pm}_{S0}$ exists are promising for switching of the large spin, but due to the fact that we deal with a multistable system, there are always additional stationary solutions present.

\textbf{Acknowledgments.} This work was supported by projects DFG BR 1528/7-1, DFG BR 1528/8-1, GRK 1558 and the Rosa Luxemburg foundation. We acknowledge fruitful discussions with C. L\'opez-Mon\'is, C. Emary, G. Kiesslich and K. Mosshammer.

\appendix

\section{Frequency doubling of $\ev{\hat S_z(t)}$} \label{App.:FrequencyDoubling}
The equation of motion for the $z$-component of the electronic spin in \Eq{eq.:SDSrateequations} does not directly couple to the magnetic field, but it couples to the product of the $x$-component for the large spin and the $y$-component of the electronic spin. The latter develop sinusoidally in time with the same frequency $\omega_s$ and without a phase shift, cf. \Fig{RegimeIIINonAd}. If we approximate their evolution in time with $\ev{\hat J^{\rm eff}_{x}(t)} = a_j \sin{\omega_s t}$ and $\ev{\hat S^{\rm eff}_{y}(t)} = a_e \sin{\omega_s t}$, we can estimate an effective solution for the $z$-component of the electronic spin ($A_s = a_j a_e$),
\begin{align}
\ev{\hat S_z^{\rm eff}(t\to \infty)} =& \frac{\left( \Gamma_{\rm L \uparrow} - \Gamma_{\rm L \downarrow} + 2 A_s  \right)}{2\Gamma}
\nonumber \\ &
- \frac{1}{8} \left[ \frac{\cos(2\omega_s t) + 2\omega_{s}/\Gamma \sin(2\omega_s t)}{(\Gamma A_s)^{-1}(\omega_s^2 - \frac{\Gamma^2}{4})}\right].
\end{align}
Hence, the oscillation goes with twice the frequency of the other spin components. Note, that for a more general ansatz, e.g.  $\ev{\hat J^{\rm eff}_{x}(t)} = a_j \sin{\omega_s t} + b_j \cos{\omega_s t}$, the result is similar and differs solely in the prefactor. 

\section{Adiabatic approach: Green's functions} \label{App.:GreenFunctions}
The expectation values of the spin operators in frequency space are obtained from
\begin{equation} \label{Eq.:AppSpinOperatorsAdiabatic}
\ev{\hat S_i (\omega)} = - \frac{i}{2} \ \rm{tr} \left[ \textbf{G}^{<} (\omega) \textbf{$\sigma$}_i \right], \hspace{0.5cm} \mbox{$i = x,y,z$},
\end{equation}
containing the Pauli spin matrices
$$\sigma_x = 
\begin{pmatrix}
 0 & 1 \\ 1 & 0 \\  
\end{pmatrix}
,
\sigma_y = 
\begin{pmatrix}
 0 & -i \\ i & 0 \\  
\end{pmatrix}
,
\sigma_z = 
\begin{pmatrix}
 1 & 0 \\ 0 & -1  \\
\end{pmatrix}
.$$
If we reconsider the effective Hamiltonian \Eq{eq.:SDShamiltonianEFF} for this system, we recall the similarity to a parallel two-level system, and as in that case, the Green's functions have matrix character,
$$\textbf{G}(\omega) = 
\begin{pmatrix}
 G_{\uparrow \uparrow }(\omega) & G_{\uparrow \downarrow}(\omega) \\ G_{\downarrow \uparrow }(\omega) & G_{\downarrow \downarrow}(\omega) \\  
\end{pmatrix}.
$$
The off-diagonal functions refer to the coupling to the $x$-component of the large spin. Without this coupling, we end up with two independent levels. 
 For the derivation of the electronic spin we need the lesser Green's function, whose calculation makes use of the Keldysh equation,
\begin{equation}
G^{<}_{\sigma \sigma'} = \sum_{\sigma''} \ G^{R}_{\sigma\sigma''}(\omega) \ \Sigma^{<}_{\sigma''\sigma''}(\omega) \ G^{A}_{\sigma'' \sigma'}(\omega),
\end{equation}
with the lesser self energy
\begin{align}
 \Sigma^{<}_{\sigma \sigma'}(\omega) = \delta_{\sigma,\sigma'} \sum_{\alpha} \Sigma^{<}_{\alpha \sigma} = 
                                         \delta_{\sigma,\sigma'} \sum_{\alpha} i \Gamma_{\alpha \sigma}  \ f_{\alpha \sigma}(\omega) .
\end{align}
The retarded and the advanced Green's function are obtained from their equation of motion and yield
\begin{align}
G_{\sigma \sigma }^{R,A}(\omega) = & \frac{\omega - \varepsilon_{\sigma'} - \Sigma^{R,A}_{\sigma'\sigma'}}{\left[\omega - \varepsilon_{\sigma} - \Sigma^{R,A}_{\sigma \sigma}\right] \hspace{-0.1cm} \left[\omega - \varepsilon_{\sigma'} - \Sigma^{R,A}_{\sigma'\sigma'}\right] \hspace{-0.05cm} -  \hspace{-0.05cm}\frac{\lambda^2}{4} \ev{\hat J_x}^2}, \nonumber \\
 G_{\sigma \sigma' }^{R,A}(\omega) = & \frac{\frac{\lambda}{2} \ev{\hat J_x}}{\omega - \varepsilon_{\sigma } - \Sigma^{R,A}_{\sigma \sigma}} \ G_{\sigma \sigma'}(\omega), 
\end{align}
with the retarded/advanced self energy 
\begin{align}
\Sigma_{\sigma \sigma'}^{R,A} = \delta_{\sigma,\sigma'} \sum_{\alpha} \Sigma_{\alpha \sigma}^{R,A}  = \delta_{\sigma,\sigma'} \sum_{k \alpha} \frac{\left|V_{k \alpha \sigma}\right|^{2}}{\omega - \varepsilon_{k \alpha \sigma} \pm i0},
\end{align}
which real part $\Lambda_{\alpha \sigma}(\omega)$ solely leads to a shift of the level energies $\varepsilon_d$ and therefore is neglected. The imaginary part corresponds to the tunneling rate $\Gamma_{\alpha \sigma}(\omega) \equiv \Gamma_{\alpha \sigma}$.

\section{Details of the stability analysis} \label{App.:FixedPoints}
The fixed points of the dynamical system are obtained from \Eq{eq.:LargeSpinEOM}, if we set all derivations to zero,
\begin{align} \label{eq.:LargeSpinEOMshort}
 0 &= -\left( \lambda  \ev{\hat S_z} + B_z \right) \  \ev{\hat J_y} , \nonumber \\
 0 &=  \hspace{0.4cm} \left( \lambda  \ev{\hat S_z} + B_z \right)  \ \ev{\hat J_x} - \lambda  \ev{\hat S_x} \ \ev{\hat J_z} ,\nonumber \\
 0 &=  \hspace{0.5cm} \lambda \ev{\hat S_x} \ \ev{\hat J_y}.
\end{align}
This is a strong nonlinear system, because the electronic spin operators depend on the large spin components $\ev{\hat S_i [\ev{\hat J_x}, \ev{\hat J_z} ] }$. Here we work in the adiabatic regime where the spin operators are given in terms of Green's functions, see \Eq{Eq.:AppSpinOperatorsAdiabatic}. The latter contain frequency integrals, leading to expressions with trigonometrical terms.
From \Eq{eq.:LargeSpinEOMshort} an analytic expression for all fixed points is not possible. Instead we obtain transcendental equations which require numerical calculations. This is the case for $\mathcal P^{\pm}_{SN}$, where the conditions $\ev{\hat S_z} =  - B_z/\lambda $ and $\ev{\hat S_x}=0$ have to be fulfilled.

The stability of the stationary solutions obtained from \Eq{eq.:LargeSpinEOMshort} are investigated with the help of the Jacobian $\mathcal J$ of the linearized system. Evaluating this Jacobian at the fixed points provides an answer if a small disturbance from these stationary solutions grows or decays, indicating unstable and stable solutions.
Here the Jacobian of the linearized dynamical system yields
\begin{widetext}
\begin{equation} \label{eq.JacobianFull}
\mathcal J = 
\begin{pmatrix} 
- \lambda \frac{\partial  \ev{\hat S_z}}{\partial \ev{\hat J_x}}  \ev{\hat J_y}
&
- (\lambda  \ev{\hat S_z} + B_z )
&
 -\lambda \frac{\partial  \ev{\hat S_z}}{\partial \ev{\hat J_z}}  \ev{\hat J_y}
\\[0.2cm]
(\lambda  \ev{\hat S_z} + B_z ) + \lambda \frac{\partial \ev{\hat S_z}}{\partial \ev{\hat J_x}} \ev{\hat J_x}
                                - \lambda \frac{\partial \ev{\hat S_x}}{\partial \ev{\hat J_x}} \ev{\hat J_z}
&
0
&
-\lambda \ev{\hat S_x} + \lambda \frac{\partial \ev{\hat S_z}}{\partial \ev{\hat J_z}} \ev{\hat J_x} 
                       - \lambda \frac{\partial \ev{\hat S_x}}{\partial \ev{\hat J_z}} \ev{\hat J_z}
\\[0.2cm]
\lambda \frac{\partial \ev{\hat S_x}}{\partial \ev{\hat J_x}} \ev{\hat J_y}
&
\lambda \ev{ \hat S_x}
&
\lambda \frac{\partial \ev{ \hat S_x} }{\partial \ev{ \hat J_z} } \ev{ \hat J_y}
\\
\end{pmatrix}.
\end{equation}
\end{widetext}
As an example we present the derivation of the fixed points $\mathcal P^{\pm}_{S0}$, where the large spin is completely polarized parallel to the magnetic field; cf. \Eq{eq.:SDSfirstFIX}. There the $x,y$-components of the electronic spin vanish due to $\ev{\hat J_{x,0}} = 0$. 
Considering the zero temperature case, where the Fermi function becomes a step function, and with $\varepsilon_{\uparrow,\downarrow}^{\pm}  = \varepsilon_d \pm 0.5 (B_z \pm j )$, we obtain for the $z$-component
\begin{align}\label{eq.:SDSspinoperatorsAd}
\ev{\hat S_{z,0}} =& \int \frac{d\omega}{4 \pi}   \sum_{\alpha}  \Theta(\mu_{\alpha} - \omega)
                     \nonumber \\ & \times
                     \left[
      \frac{ \Gamma_{ \alpha \uparrow}}
           {\left| \left[\omega - \varepsilon_{\uparrow }^{\pm}   + i \frac{\Gamma}{2}   \right]   \right|^2} 
    - \frac{ \Gamma_{\alpha \downarrow} }
           {\left|  \left[\omega - \varepsilon_{\downarrow }^{\pm}  + i \frac{\Gamma}{2}\right] \right|^2} \right]
            \nonumber \\
             =& \sum_{\alpha} \left\{ \frac{\Gamma_{\alpha\uparrow}}{2\pi\Gamma} \arctan \left[ \frac{(\mu_{\alpha} -\varepsilon_{\uparrow}^{\pm}  )}{\Gamma/2}\right] \right. \nonumber \\ & \left. \hspace{0.5cm}
                                        - \frac{\Gamma_{\alpha\downarrow}}{2\pi\Gamma} \arctan\left[ \frac{(\mu_{\alpha} -\varepsilon_{\downarrow}^{\pm} )}{\Gamma/2}\right] \hspace{-0.1cm}
                                        +\frac{\left( \Gamma_{ \alpha \uparrow} - \Gamma_{ \alpha \downarrow} \right)}{4\Gamma} \right\}.
\end{align}
Due to $\Gamma = \Gamma_{\sigma \rm{L}} +  \Gamma_{\sigma \rm{R}}$ the last term in \Eq{eq.:SDSspinoperatorsAd} vanishes if the summation over $\alpha$ is accomplished and the result \Eq{eq.:SzFirstFIX} is obtained. 

The Jacobian simplifies to 
\begin{equation} \label{eq.AppJacobiFirstFIX}
\mathcal J^{\pm}_{0} = 
\begin{pmatrix} 
0
&
\hspace{-0.5cm} - (\lambda  \ev{\hat S_{z,0}} + B_z)
&
0
\\[0.2cm]
  \lambda  \ev{\hat S_{z,0}} + B_z \mp \left. \lambda j \frac{\partial \ev{\hat S_x}}{\partial \ev{\hat J_x}} \right|_{\substack{\mathcal P^{\pm}_{S0}} } 
&
0
&
0
\\[0.2cm]
0
&
0
&
0
\\
\end{pmatrix},
\end{equation}
from which the eigenvalues $\mathcal E_{2,3}^{0,\pm} $ are calculated, cf. \Eq{eq.:SDSeigenvaluesFIX1adiabatic}.
The derivation in \Eq{eq.AppJacobiFirstFIX} is obtained from
\begin{align}
 \left. \frac{\partial \ev{\hat S_x}}{\partial \ev{\hat J_x}} \right|_{\substack{\mathcal P^{\pm}_{S0}} }  =& \
 \lambda  \int  \frac{d\omega}{4 \pi} \sum_{\alpha} \Theta(\mu_{\alpha} - \omega) \nonumber \\ & \times
      \frac{\left[ (\omega - \varepsilon_{\downarrow}^{\pm}) \Gamma_{\alpha \uparrow}
                                       + (\omega - \varepsilon_{\uparrow}^{\pm}) \Gamma_{\alpha \downarrow}  \right]}
           {\left| \left[\omega - \varepsilon_{\uparrow }^{\pm} + i \frac{\Gamma}{2}\right]
                   \left[\omega - \varepsilon_{\downarrow }^{\pm} + i \frac{\Gamma}{2}\right] \right|^2}.
\end{align}

\end{document}